\documentclass[aps,prb,twocolumn,floatfix, 10pt, superscriptaddress,eqsecnum,showpacs,showkeys,amsmath,amssymb]{revtex4}
\usepackage{amsfonts}
\usepackage{amssymb,amsmath}
\usepackage{graphicx}
\usepackage{dcolumn}
\usepackage{bm}
\usepackage{color}
\def\2{\frac{1}{2}} \def\4{\frac{1}{4}}
\def\6{\partial}
\def\+{\dagger}
\def\<{\langle} \def\>{\rangle}

\def\i{{\rm i}}

\DeclareMathOperator{\sh}{sh}
\renewcommand{\sinh}{\sh}
\DeclareMathOperator{\ch}{ch}
\renewcommand{\cosh}{\ch}

\renewcommand{\tilde}{\widetilde}

\begin{document}
\title{
Correlation functions in one-dimensional spin lattices with Ising
and Heisenberg bonds}

\author{Stefano Bellucci}
\affiliation{INFN-Laboratori Nazionali di Frascati, Via E. Fermi 40,
00044 Frascati, Italy}
\author{Vadim Ohanyan}
\affiliation{Department of Theoretical Physics,
             Yerevan State University,
             Alex Manoogian 1, 0025 Yerevan, Armenia}

\affiliation{ICTP, Strada Costiera 11, I-34151 Trieste, Italy}

\date{\today}

\begin{abstract}
A general technique of exact calculation of any correlation
functions for the special class of one-dimensional spin models
containing small clusters of quantum spins assembled to a chain by
alternating with the single Ising spins is proposed. The technique
is a natural generalization of that in the models solved by a
classical transfer matrix. The general expressions for corresponding
matrix operators which are the key components of the
technique are obtained. 
As it is clear from the general principles, the decay of the
correlation functions of various types is explicitly shown to be
     governed by a single correlation length.
The technique is illustrated by two examples: symmetric diamond
chain and asymmetric sawtooth chain.
\end{abstract}



\maketitle

\section{Introduction}

  Recently, much attention has been paid to the exactly solvable models
of one-dimensional magnetism, consisting of small quantum spin
clusters connected to each other via Ising
spins.\cite{str03,str04,str05,can06,val08,per08,per09,str09,ant09,oha09,oha10,
str10,bel10,roj11a,roj11b,lis11a,lis11b,str11,oha12,cha12,roj12a,ana12,ver12,str12,gal,roj12b}
For the sake of simplicity we will call them Heisenberg--Ising
chains (HIC). The interest toward such spin systems stems from
several features they possess. First of all, they allow one to
obtain an exact statistical-mechanical solution of the problem of
strongly interacting spins in terms of classical transfer
matrix\cite{bax, oha03,ayd05,ayd06}. In addition to that, such
systems combine the quantum and classical properties and also, to
some extent, are connected to the problems and models of molecular
magnetism\cite{mol_mag}, as the classical transfer matrix is
constructed by the diagonalization of the small quantum spin
cluster. And finally, very recently the family of trimetallic
coordination polymer compounds has been synthesized\cite{chem, ring,
Dy_10}. These compounds in the magnetic sense are indeed a
one-dimensional magnets with both classical (Ising) and quantum
(Heisenberg) bonds. The single-chain magnet reported in Ref.\
\onlinecite{chem} is an example of a Heisenberg-Ising chain of
triangles with three-spin linear quantum clusters; another
interesting molecular magnet (spin ring) based on the highly
anisotropic properties of the Dy$^{3+}$ ions has been reported in
Ref.\ \onlinecite{ring}. Both systems have been analyzed by the
transfer-matrix solution of HIC chains in Ref.\ \onlinecite{Dy_10}.
Another class of examples is given by chain magnets with two
alternating magnetic ions, thus giving rise to alternating quantum
and classical spins (see, e.g.,
Refs.~\onlinecite{RCRS04,JACS10,SSR11}).
 Let us describe the general scheme of the HIC:\\
\begin{itemize}

  \item There is a
 small cluster of a few quantum spins with a certain topology of bonds
which could contain various types of spin-spin interaction as well
as interactions with external fields. The Hamiltonian of the single
small cluster is, thus, a finite dimensional matrix. \\

\item These clusters are assembled to the chain or any other
one-dimensional structure by alternating with the single Ising spins.\\

\item The interaction between the single Ising spins and any quantum
spin from a small cluster must contain only $z$-components (Ising
interaction).

\end{itemize}

For the mixed quantum--classical spin system with the above
mentioned properties the exact solution is obtained in the following
way. The idea is to apply a classical transfer-matrix formalism in
which the entries of the classical transfer matrix are obtained from
the diagonalization of a small quantum spin cluster. Although the
number of papers devoted to the exact solutions of the HIC and to
their magneto-thermal properties is rather impressive, only a few
models have been solved within the classical transfer matrix
formalism\cite{val08,ant09,oha09, oha10, bel10, roj11a, roj11b,
lis11a,lis11b, oha12,cha12,roj12a,ana12,roj12b}, while the other
method of the so-called decoration iteration transformation has been
also used\cite{str03,str04,str05,can06,str09,str10,str12,gal}.
Particularly, within the direct classical transfer matrix technique
the following models have been solved exactly: the
double-tetrahedral chain with quantum triangles of
spins\cite{ant09,oha10}, the sawtooth chain with an asymmetric pair
of quantum spins\cite{oha09, bel10}, diamond
chain\cite{roj11a,lis11a,lis11b,cha12,roj12a,ana12,roj12b} and
orthogonal-dimer chain with anisotropic and asymmetric quantum
triangles\cite{oha12}.
 Despite the complete and exact description of the ground states and
 ground states phase diagrams, as well of all thermodynamic functions,
 finite-temperature entropy and magnetization, magnetocaloric
 effect, etc, the expressions for the correlation function are still
 missing. It is worth mentioning, that the correlation functions can be in principle calculated
 by another method which is used to handle the HIC, the technique of the decoration iteration
 transformation\cite{str03,str04,str05,can06,str09,str10,str12,gal,fisher,roj09,str10c}. However, the method of the classical transfer
 matrix
 is more straightforward and is based on a direct application of the canonical expectation
 values\cite{kne,lyra}, which can be calculated exactly.
    In the present paper we are going to develop an exact technique for the calculation of any finite-temperature correlation
 function for any HIC by means of the direct application of the classical transfer-matrix method.

  Beside the pure academic interest, the knowledge about the exact
  correlation functions of the HIC recently became important from
  the experimental point of view as well. As mentioned above, a few years ago it was
  reported about the synthesis of a novel class of trimetallic
  coordination polymer compounds\cite{chem, ring, Dy_10, JACS10,
  SSR11} which are the real examples of the HIC (at least one of
  them). As known, the neutron scattering is one of the most
  powerful and straightforward methods of investigation of the
  magnetic structure, spin textures, magnetic excitations,
  etc. In the leading order the cross-section of the neutron
  scattering is given by the two-point spin-spin
  correlation functions (their Fourier-transform). This fact makes
  the topic of the present paper a practical issue.

    The paper is organized as follows. In the second Section we remind
  the reader of the main features of the classical transfer-matrix
  formalism for the HIC, in the next section we present the
  formalism for calculating the correlation function of HIC in terms
  of traces of certain matrix products and prove that for all kinds of correlation functions the correlation length is the same.
   In Section IV we apply
  the general formalism for several models. The paper ends with a
  Conclusion.

  \section{Classical transfer-matrix formalism for HIC}
The Hamiltonian of a HIC has the form of a sum of  local block
Hamiltonians ${\mathcal{H}}_n$ which correspond to the n-th small
quantum cluster and its interaction with left and right Ising spins:
\begin{equation}\label{ham}
{\mathcal{H}}=\sum_{n=1}^N\left( {\mathcal{H}}_n-\frac{H}{2}
\left(\sigma_n+\sigma_{n+1} \right)\right),
\end{equation}
here $\sigma_n$ is the Ising spin situated between the n-th and
(n+1)-th small cluster and $H$ is the magnetic field acting on Ising
spins. The block Hamiltonian can contain various terms with quantum
and classical spin variables which are determined by the topology of
the elementary block and the character of interactions. Let us
denote the quantum spins by $\mathbf{S}$; then one can suppose that
the block Hamiltonian depends on $\mathbf{S}$, $\sigma$ and may be
on other classical spins $\tau_b$,
\begin{eqnarray}
{\mathcal{H}}_n={\mathcal{H}}\left({\mathbf{S}}_{n,a}, \sigma_n,
\sigma_{n+1}, \tau_{n,b}\right). \label{block_ham}
\end{eqnarray}
Due to the properties listed above the block Hamiltonians commute to
each other,
\begin{equation}\label{com}
\left[{\mathcal{H}}_i, {\mathcal{H}}_j \right]=0.
\end{equation}
This fact allows one to expand the exponential in the expression for
the partition function of the model and obtain an expression which
by its structure corresponds to the classical transfer matrix
scheme:
\begin{widetext}
\begin{eqnarray}
{\mathcal{Z}}&=&{\mathrm{Tr}}_{{\mathrm{S}}}\sum_{(\sigma)}\sum_{(\tau)}
\exp\left(-\beta {\mathcal{H}}\right)=\sum_{(\sigma)}\prod_{n=1}^N
e^{\beta \frac{H}{2}\left(\sigma_n+\sigma_{n+1}
\right)}\sum_{(\tau)} {\mathrm{Tr}}_n
e^{-\beta{\mathcal{H}}_n}=\\
&&\sum_{(\sigma)}\prod_{n=1}^N e^{\beta
\frac{H}{2}\left(\sigma_n+\sigma_{n+1}
\right)}\sum_{(\tau)}\sum_{i=1}^L e^{-\beta \lambda_i\left(\sigma_n,
\sigma_{n+1}|\tau_{n,b} \right)}=\sum_{(\sigma)}\prod_{n=1}^N
T_{\sigma_n, \sigma_{n+1}}={\mathrm{Tr}}\mathbf{T}^N, \nonumber
\label{Z2}
\end{eqnarray}
\end{widetext}
here the cyclic boundary conditions are supposed,
$\lambda_i(\sigma_n,\sigma_{n+1}|\tau_{n,b})$ are the eigenvalues of
the block Hamiltonian ${\mathcal{H}}_n$(which is supposed to be an
$L$ by $L$ matrix ) depending on the values of classical spins
$\tau$ and classical spins $\sigma_n$ which alternate with the
blocks, ${\mathrm{Tr}}_n$ means the trace over the states of the
$n$-th quantum block, and ${\mathrm{Tr}}$ stands for the trace of
the 2 by 2 transfer matrix with entries $T_{\sigma_n,
\sigma_{n+1}}$. After summation over the values of $\tau$ for each
block one gets the expression depending only on $\sigma_n$ and
$\sigma_{n+1}$ which is the classical 2 by 2 transfer matrix
\begin{eqnarray}
\mathbf{T}=\left( \begin{array}{lcr}
      e^{\beta \frac{H}{2}} Z_{++}  & Z_{+-} \\
      Z_{-+}  &  e^{-\beta \frac{H}{2}} Z_{--} \label{T}
      \end{array}\right),
\end{eqnarray}
where $Z_{\sigma_n, \sigma_{n+1}}=\sum_{(\tau)}\sum_{i=1}^L
e^{-\beta \lambda_i\left(\sigma_n, \sigma_{n+1}|\tau_{n,b} \right)}$
is the partition function of one block, depending on the values of
two neighboring Ising spins, signs "$+$" and "$-$" in indexes
correspond to $\sigma=1/2$ and $\sigma=-1/2$ respectively.
The formalism of classical transfer-matrix allows one to
obtain exact analytic expressions for almost any expectation values
of the function of local variables (spins) and, of course, for any
thermodynamic functions, as the expression for the free energy per
block can be obtained in a straightforward way by finding the
maximal eigenvalue of the transfer matrix (\ref{T}):
\begin{widetext}
\begin{eqnarray}
f=-\frac{1}{\beta}\log\frac{1}{2}\left(e^{\beta
\frac{H}{2}}Z_{++}+e^{-\beta \frac{H}{2}}
Z_{--}+\sqrt{\left(e^{\beta \frac{H}{2}}Z_{++}-e^{-\beta
\frac{H}{2}} Z_{--} \right)^2+4 Z_{+-}Z_{-+}} \right).
\label{free_en}
\end{eqnarray}
\end{widetext}
From this expression any thermodynamic function, like magnetization,
entropy, specific head etc can be obtained exactly. However, to
obtain correlation functions of the spins demands more efforts and
further development of the classical transfer-matrix technique.
\section{Correlation functions and correlation length}
 Let us concentrate on the scheme of calculation of the expectation values of combinations of spins operators.
 First of all, let us start from the simplest case, the expectation values of $\sigma$-s, which is quite similar to that of the Ising chain\cite{bax}.
 Thus, for the thermodynamic average of just one $\sigma$ one
 obtains
\begin{eqnarray}
\langle \sigma_j
\rangle&&=\frac{1}{{\mathcal{Z}}}\sum_{(\sigma)}\sigma_j\prod_{n=1}^N
T_{\sigma_n, \sigma_{n+1}}\nonumber
\\
&&=\frac{1}{{\mathcal{Z}}}\sum_{(\sigma)}T_{\sigma_1,
\sigma_{2}}...\sigma_j T_{\sigma_j, \sigma_{j+1}}...T_{\sigma_N,
\sigma_{1}}. \label{M_sig}
\end{eqnarray}
This expression is the trace of the product of $N$ transfer-matrices
with the matrix
\begin{eqnarray}
{\mathbf{\sigma}}^z=\left( \begin{array}{lcr}
      1/2  & 0 \\
      0  &  -1/2 \label{sigm_z}
      \end{array}\right)
\end{eqnarray}
inserted before the $j$-th transfer matrix in the trace formula.
Thus, we have
\begin{eqnarray}
\langle \sigma_j
\rangle=\frac{1}{{\mathcal{Z}}}{\mathbf{Tr}}\left({\mathbf{\sigma}}^z{\mathbf{T}}^N\right)=\frac{1}{{\mathcal{Z}}}{\mathbf{Tr}}
\left({\mathbf{\sigma}}^z{\mathbf{A}}{\mathbf{\Lambda}}^N{\mathbf{A}}^{-1}\right),
\label{M_sig2}
\end{eqnarray}
where we used the matrix identity
${\mathbf{T}}={\mathbf{A}}{\mathbf{\Lambda}}{\mathbf{A}}^{-1}$, with
the diagonal matrix ${\mathbf{\Lambda}}$ and the orthogonal matrix
${\mathbf{A}}$. Thus,
\begin{eqnarray}
{\mathbf{\Lambda}}=\left( \begin{array}{lcr}
      \lambda_1  & 0 \\
      0  &  \lambda_2 \label{Lambda}
      \end{array}\right), \qquad
{\mathbf{A}}=\left( \begin{array}{lcr}
      \cos \phi  & -\sin \phi \\
      \sin \phi  &  \cos\phi
      \end{array}\right),
\end{eqnarray}
where $\lambda_{1,2}$ are the eigenvalues of the transfer-matrix,
and $\phi$ is a parameter. Having all these notations we now can
express the canonical expectation values from Eq. (\ref{M_sig}) in
terms of the entries of the orthogonal matrix diagonalizing the
transfer-matrix. Using the properties of the trace (cyclic
permutation inside the trace does not change it) one can get
\begin{eqnarray}
\langle \sigma_j \rangle&=&\frac{1}{{\mathcal{Z}}}{\mathbf{Tr}}
\left({\mathbf{A}}^{-1}{\mathbf{\sigma}}^z{\mathbf{A}}{\mathbf{\Lambda}}^N\right)=\nonumber
\\&&\frac{1}{2}\frac{(\cos2\phi)
\lambda_1^2-(\sin2\phi)\lambda_2^N}{\lambda_1^N+\lambda_2^N}
\label{M_sig3}
\end{eqnarray}
Now, we should take care about the thermodynamic limit
($N\rightarrow\infty$). It is easy to see, that in this limit Eq.
(\ref{M_sig3}) leads to (suppose that $\lambda_1$ is greater than
$\lambda_2$)
\begin{eqnarray}
\langle \sigma_j \rangle=\frac{1}{2}\cos2\phi\equiv
\frac{1}{2}M_{\sigma},\label{M_sig4}
\end{eqnarray}
where $M_{\sigma}$ is the partial magnetization corresponding to the
subsystem with $\sigma$-spins,
\begin{eqnarray}\label{M}
M_{\sigma}=\frac{e^{\beta \frac{H}{2}}Z_{++}-e^{-\beta
\frac{H}{2}}Z_{--}}{\sqrt{\left(e^{\beta
\frac{H}{2}}Z_{++}-e^{-\beta \frac{H}{2}}Z_{--} \right)^2+4
Z_{+-}Z_{-+}}} \label{M_via_T}
\end{eqnarray}
 Thus, we can express the entries of
the orthogonal matrix diagonalizing the transfer matrix in terms of
the subsystem magnetization $M_{\sigma}$:
\begin{eqnarray}
{\mathbf{A}}=\frac{1}{\sqrt{2}}\left(
\begin{array}{lcr}
      \sqrt{1+ M_{\sigma}}  & -\sqrt{1- M_{\sigma}} \\
       \sqrt{1- M_{\sigma}} &  \sqrt{1+ M_{\sigma}} \label{A}
      \end{array}\right)
\end{eqnarray}
These quantities are the building blocks for the calculation of all
correlation function for the HIC. As will be clear from further
consideration, any correlation function of HIC has the form of the
trace of a certain product of transfer matrices and special matrices
representing the spin operators. Moreover, these matrices are
constructed from the local expectation values of spin operators with
the aid of the matrix $\mathbf{A}$ in the standard way
$\tilde{\mathbf{P}}\equiv\mathbf{A}^{-1}\mathbf{P}\mathbf{A}$. Let
us write down the entries of a general matrix of that type
\begin{widetext}
\begin{eqnarray}\label{tild_P_ent}
&&\tilde{P}_{++}=\frac{1}{2}\left[\left(1+M_{\sigma}\right)P_{++}+\sqrt{1-M_{\sigma}^2}\left(
P_{+-}+P_{-+}\right)+ \left(1-M_{\sigma}\right)P_{--}\right], \\
&&\tilde{P}_{+-}=\frac{1}{2}\left[\sqrt{1-M_{\sigma}^2}\left(P_{--}-P_{++}\right)
+\left(1+M_{\sigma}\right)P_{+-}-\left(1-M_{\sigma}\right)P_{-+}\right],\nonumber \\
&&\tilde{P}_{-+}=\frac{1}{2}\left[\sqrt{1-M_{\sigma}^2}\left(P_{--}-P_{++}\right)
+\left(1+M_{\sigma}\right)P_{-+}-\left(1-M_{\sigma}\right)P_{+-}\right],\nonumber \\
&&\tilde{P}_{--}=\frac{1}{2}\left[\left(1-M_{\sigma}\right)P_{++}-\sqrt{1-M_{\sigma}^2}\left(
P_{+-}+P_{-+}\right)+
\left(1+M_{\sigma}\right)P_{--}\right].\nonumber
\end{eqnarray}
\end{widetext}
For instance, the matrix
${\mathbf{\widetilde{\sigma}}}^z={\mathbf{A}}^{-1}{\mathbf{\sigma}}_z{\mathbf{A}}$
which is the building block for constructing various correlation
functions of $\sigma_j$ has a particulary simple form:
\begin{eqnarray}
{\mathbf{\widetilde{\sigma}}}^z=\frac{1}{2}\left(
\begin{array}{lcr}
       M_{\sigma}  & -\sqrt{1- M_{\sigma}^2} \\
       -\sqrt{1- M_{\sigma}^2} &  - M_{\sigma} \label{tild_sigma}
      \end{array}\right)
\end{eqnarray}
 The simple two-site correlation function is just a thermal
average of the product of two $\sigma$ spins situated at a distance
$r$ from each other:
\begin{widetext}
\begin{eqnarray}
\langle \sigma_j
\sigma_{j+r}\rangle=\frac{1}{{\mathcal{Z}}}\sum_{(\sigma)}\sigma_j
\sigma_{j+r} \prod_{n=1}^N T_{\sigma_n,
\sigma_{n+1}}=\frac{1}{{\mathcal{Z}}}{\mathbf{Tr}}\left({\mathbf{T}}^{N-r}{\mathbf{\sigma}}^z{\mathbf{T}}^r{\mathbf{\sigma}}^z\right)=
\frac{1}{{\mathcal{Z}}}{\mathbf{Tr}}\left({\mathbf{\Lambda}}^{N-r}{\mathbf{\widetilde{\sigma}}}^z
{\mathbf{\Lambda}}^r{\mathbf{\widetilde{\sigma}}}^z\right).
\label{cor_sigma_1} \nonumber
\end{eqnarray}
\end{widetext}
Thus, using the matrix form of ${\mathbf{\widetilde{\sigma}}}^z$ and
taking a thermodynamic limit one obtains:
\begin{eqnarray}\label{cor_sig_sig}
&&\langle \sigma_j
\sigma_{j+r}\rangle=\frac{1}{4}\left(M_{\sigma}^2+\left(1-M_{\sigma}^2
\right)e^{-\frac{r}{\xi}}\right), \\
&&\xi=\frac{1}{\log \left(\frac{\lambda_1}{\lambda_2}\right)},
\nonumber \label{cor_sigma_3_xi}
\end{eqnarray}
where $\xi$ is a correlation length. So far, all results have the
same form as those for a simple Ising chain, the difference is only
in the form of the magnetization for $\sigma$-spins, which in our
case is given by Eqs. (\ref{M_via_T}) and (\ref{T}).  Now, let us
turn to the more complicated correlation functions of the quantum
spins with each other and mixed correlation functions,
\begin{widetext}
\begin{eqnarray}\label{cor_SS_Ssig}
&&\langle S_{j,a}^{\alpha}
S_{j+r,b}^{\beta}\rangle=\frac{1}{{\mathcal{Z}}}{\mathbf{Tr}}_S\sum_{(\sigma)}\sum_{(\tau)}S_{j,a}^{\alpha}
S_{j+r,b}^{\beta}\prod_{n=1}^N
e^{-\beta{{\mathcal{H}}}_n}e^{\beta\frac{H}{2}\left(\sigma_n+\sigma_{n+1}\right)}
=\frac{1}{{\mathcal{Z}}}{{\mathbf{Tr}}}\left({\mathbf{\Lambda}}^{N-r-1}{\tilde{\mathbf{\Sigma}}}_a^{\alpha}{{\mathbf{\Lambda}}}^{r-1}
{\tilde{\mathbf{\Sigma}}}_b^{\beta} \right), \\
&&\langle S_{j,a}^{\alpha}
\sigma_{j+r}\rangle=\frac{1}{{\mathcal{Z}}}{\mathbf{Tr}}_S\sum_{(\sigma)}\sum_{(\tau)}S_{j,a}^{\alpha}
\sigma_{j+r}\prod_{n=1}^N e^{-\beta
{{\mathcal{H}}}_n}e^{\beta\frac{H}{2}\left(
\sigma_n+\sigma_{n+1}\right)}=\frac{1}{{\mathcal{Z}}}{{\mathbf{Tr}}}\left({\mathbf{\Lambda}}^{N-r-1}{\tilde{\mathbf{\Sigma}}}_a^{\alpha}{{\mathbf{\Lambda}}}^{r}
{\tilde{\mathbf{\sigma}}}^{z} \right) \nonumber\\
&&\langle S_{j,a}^{\alpha}
\tau_{j+r,b}\rangle=\frac{1}{{\mathcal{Z}}}{\mathbf{Tr}}_S\sum_{(\sigma)}\sum_{(\tau)}S_{j,a}^{\alpha}
\tau_{j+r,b}\prod_{n=1}^N e^{-\beta
{{\mathcal{H}}}_n}e^{\beta\frac{H}{2}\left(
\sigma_n+\sigma_{n+1}\right)}=\frac{1}{{\mathcal{Z}}}{{\mathbf{Tr}}}\left({\mathbf{\Lambda}}^{N-r-1}{\tilde{\mathbf{\Sigma}}}_a^{\alpha}{{\mathbf{\Lambda}}}^{r-1}
{\tilde{\mathbf{\tau}}}_b \right),\nonumber\\
&&\langle \tau_{j,a} S_{j+r,b}^{\alpha}
\rangle=\frac{1}{{\mathcal{Z}}}{\mathbf{Tr}}_S\sum_{(\sigma)}\sum_{(\tau)}
\tau_{j,a}S_{j+r,b}^{\alpha}\prod_{n=1}^N e^{-\beta
{{\mathcal{H}}}_n}e^{\beta\frac{H}{2}\left(
\sigma_n+\sigma_{n+1}\right)}=\frac{1}{{\mathcal{Z}}}{{\mathbf{Tr}}}\left({\mathbf{\Lambda}}^{N-r-1}{\tilde{\mathbf{\tau}}}_a{{\mathbf{\Lambda}}}^{r-1}
{\tilde{\mathbf{\Sigma}}}_b^{\alpha} \right),\nonumber\\
 &&\langle\tau_{j,a}
\tau_{j+r,b}\rangle=\frac{1}{{\mathcal{Z}}}{\mathbf{Tr}}_S\sum_{(\sigma)}\sum_{(\tau)}\tau_{j,a}
\tau_{j+r,b}\prod_{n=1}^N e^{-\beta
{{\mathcal{H}}}_n}e^{\beta\frac{H}{2}\left(
\sigma_n+\sigma_{n+1}\right)}=\frac{1}{{\mathcal{Z}}}{{\mathbf{Tr}}}\left({\mathbf{\Lambda}}^{N-r-1}{\tilde{\mathbf{\tau}}}_a{{\mathbf{\Lambda}}}^{r-1}
{\tilde{\mathbf{\tau}}}_b \right),\nonumber\\
&&\langle \tau_{j,a}
\sigma_{j+r}\rangle=\frac{1}{{\mathcal{Z}}}{\mathbf{Tr}}_S\sum_{(\sigma)}\sum_{(\tau)}\tau_{j,a}
\sigma_{j+r}\prod_{n=1}^N e^{-\beta
{{\mathcal{H}}}_n}e^{\beta\frac{H}{2}\left(
\sigma_n+\sigma_{n+1}\right)}=\frac{1}{{\mathcal{Z}}}{{\mathbf{Tr}}}\left({\mathbf{\Lambda}}^{N-r-1}{\tilde{\mathbf{\tau}}}_a{{\mathbf{\Lambda}}}^{r}
{\tilde{\mathbf{\sigma}}} \right),\nonumber\\
&&\langle \sigma_{j} S_{j+r,a}^{\alpha}
\rangle=\frac{1}{{\mathcal{Z}}}{\mathbf{Tr}}_S\sum_{(\sigma)}\sum_{(\tau)}\sigma_{j}S_{j+r,a}^{\alpha}
\prod_{n=1}^N e^{-\beta {{\mathcal{H}}}_n}e^{\beta\frac{H}{2}\left(
\sigma_n+\sigma_{n+1}\right)}=\frac{1}{{\mathcal{Z}}}{{\mathbf{Tr}}}\left({{\mathbf{\Lambda}}}^{N-r}
{\tilde{\mathbf{\sigma}}}^{z}{\mathbf{\Lambda}}^{r-1}{\tilde{\mathbf{\Sigma}}}_a^{\alpha}
\right) \nonumber \\
&&\langle \sigma_{j}\tau_{j+r,a}
\rangle=\frac{1}{{\mathcal{Z}}}{\mathbf{Tr}}_S\sum_{(\sigma)}\sum_{(\tau)}\sigma_{j}\tau_{j+r,a}
\prod_{n=1}^N e^{-\beta {{\mathcal{H}}}_n}e^{\beta\frac{H}{2}\left(
\sigma_n+\sigma_{n+1}\right)}=\frac{1}{{\mathcal{Z}}}{{\mathbf{Tr}}}\left({{\mathbf{\Lambda}}}^{N-r}
{\tilde{\mathbf{\sigma}}}{\mathbf{\Lambda}}^{r-1}{\tilde{\mathbf{\tau}}}_a
\right),\nonumber
\end{eqnarray}
\end{widetext}
 where the new 2 by 2 matrices ${\mathbf{\Sigma}}_a^{\alpha}$ and $\mathbf{\tau}_a$ are the classical transfer-matrix formalism
 representation of
 the $\alpha$-component of the $a$-th quantum spin and classical $a$-th spins from the
 corresponding quantum clusters. Their entries are found according to
\begin{eqnarray}
&&\left(\Sigma_{a}^{\alpha}\right)_{\sigma_j,
\sigma_{j+1}}=e^{\beta\frac{H}{2}\left( \sigma_j+
\sigma_{j+1}\right)}\sum_{(\tau_j)}{\mathbf{Tr}}_j\left(S_{j,a}^{\alpha}e^{-\beta
{\mathcal{H}}_j} \right) \nonumber \\
&&\left(\tau_{a}\right)_{\sigma_j,
\sigma_{j+1}}=e^{\beta\frac{H}{2}\left( \sigma_j+
\sigma_{j+1}\right)}\sum_{(\tau_j)}\tau_{j,a}{\mathbf{Tr}}_j
e^{-\beta {\mathcal{H}}_j}
 \label{Sigma,tau}
\end{eqnarray}
The correlation functions containing the quantum spins $S$ have the
unusual overall power of the ${\mathbf{\Lambda}}$-matrix smaller
than the number of the blocks $N$. The point is that the inclusion
of each quantum spin into the function reduces the overall power of
the ${\mathbf{\Lambda}}$-matrix by one. This technical difference
from the standard classical technique\cite{kne,lyra} stems out from
the following feature. One does not have the spin operators
${\mathbf{S}}$ by themselves under the trace in the canonical
expectation value expression corresponding to the correlation
function. Instead, there are some complicated matrices given by Eq.
(\ref{Sigma,tau}) and representing the quantum spin operators in the
generalized classical transfer-matrix formalism. As it is seen from
Eq. (\ref{Sigma,tau}), due to the quantum nature of the variables
$S$, the corresponding matrix representation for them adsorbs the
transfer matrix and it is an issue to find a representation where
the resulting matrix under the trace into the partition function
would be just a product of the transfer-matrix and some other
matrix. On the contrary, for the classical spin chains for arbitrary
spin the expression $S_jT_{S_j,S_{j+1}}$ is the analog of the Eq.
(\ref{Sigma,tau}) which appears under the trace formula in the
corresponding correlation function. Thus, the corresponding matrix
to be inserted into the trace formula can be easily and naturally
split into the product of the transfer matrix and a diagonal matrix
with the possible values of the classical variable $S$ on its
diagonal\cite{bax, kne, lyra}. This fact leads also to another
consequence concerning the properties of the correlation functions
from Eq. (\ref{cor_SS_Ssig}). The expressions are no longer valid
for $r=0$. All these cases should be considered separately out of
the framework of the technique developed here. However, the
correlation functions at $r=0$ for the same spin operator are
trivial because they are just a expectation values of the square of
spin, while correlation functions for the different spins belonging
to the same block can be easily calculated by taking a derivatives
of the free energy with respect to the corresponding coupling
constants.

 As the expressions
under the trace in Eq.(\ref{cor_SS_Ssig}) have the same structure,
let us write down a general expression for such a correlation
function
\begin{eqnarray}\label{CorPR}
&&\frac{1}{{\mathcal{Z}}}{{\mathbf{Tr}}}\left({\mathbf{\Lambda}}^{N-r-1}{\tilde{\mathbf{P}}}{{\mathbf{\Lambda}}}^{r-1}
{\tilde{\mathbf{R}}}
\right)\\
&&=\frac{1}{\lambda_1^2}\tilde{P}_{++}\tilde{R}_{++}+\frac{1}{\lambda_1\lambda_2}\tilde{P}_{+-}\tilde{R}_{-+}
e^{-r/\xi},\nonumber
\end{eqnarray}
where the correlation length $\xi$ is the same as in
Eq.(\ref{cor_sig_sig}). Obviously, the part independent of $r$ is
the product of expectation values of quantities corresponding to
matrices $\mathbf{P}$ and $\mathbf{R}$,
\begin{eqnarray}\label{expP}
\langle P_j
\rangle=\frac{1}{{\mathcal{Z}}}{{\mathbf{Tr}}}\left({\mathbf{\Lambda}}^{N-1}{\tilde{\mathbf{P}}}\right)=\frac{1}{\lambda_1}\tilde{P}_{++}.
\end{eqnarray}
The product of two eigenvalues of the 2 by 2 transfer matrix is its
determinant, thus, eventually one can write down the following
general form of pair-correlation function for HIC:
\begin{eqnarray}\label{cor_PR2}
&&\langle P_j
R_{j+r}\rangle=\frac{1}{{\mathcal{Z}}}{{\mathbf{Tr}}}\left({\mathbf{\Lambda}}^{N-r-1}{\tilde{\mathbf{P}}}{{\mathbf{\Lambda}}}^{r-1}
{\tilde{\mathbf{R}}}
\right)\\
&&=\langle P_j \rangle\langle R_j\rangle+\frac{\tilde{P}_{+-}
\tilde{R}_{-+}}{\det \mathbf{T}}e^{-r/\xi}\nonumber
\end{eqnarray}
 If one of
the matrices is $\sigma$ the result is different, because the
corresponding expression contains one more transfer matrix. In
general, it also depends on the position of the $\sigma$, unless the
matrix ${\tilde{\mathbf{P}}}$ is symmetric. Thus, the corresponding
expressions for the correlation functions are
\begin{eqnarray}\label{corr_Psig}
&&\frac{1}{{\mathcal{Z}}}{{\mathbf{Tr}}}\left({\mathbf{\Lambda}}^{N-r-1}{\tilde{\mathbf{P}}}{{\mathbf{\Lambda}}}^{r}
{\tilde{\mathbf{\sigma}}}
\right)\\
&&=\langle P_j\rangle\langle
\sigma_j\rangle-\frac{\tilde{P}_{+-}\sqrt{1-M_{\sigma}^2}}{2\lambda_1}
e^{-r/\xi},\nonumber\\
&&\frac{1}{{\mathcal{Z}}}{{\mathbf{Tr}}}\left({\mathbf{\Lambda}}^{N-r}{\tilde{\mathbf{\sigma}}}{{\mathbf{\Lambda}}}^{r-1}{\tilde{\mathbf{P}}}
\right)\nonumber\\
&&=\langle P_j\rangle\langle
\sigma_j\rangle-\frac{\tilde{P}_{-+}\sqrt{1-M_{\sigma}^2}}{2\lambda_1}
e^{-r/\xi}.\nonumber
\end{eqnarray}
The matrix ${\tilde{\mathbf{P}}}$, in its turn, is symmetric when
the corresponding matrix $\mathbf{P}$ is symmetric. Thus, the order
of $P_j$ and $\sigma_{i}$ in the correlation function does not
matter in this case: $\langle P_j
\sigma_{j+r}\rangle=\langle\sigma_j P_{j+r}\rangle$.
 From this general consideration one can see that there exists only a single correlation
 length for any kind of correlation functions of the HIC.
This is rather expected result because the appearance of a single
correlation length even in the inhomogeneous situation is the common
feature of all one-dimensional systems which do not become critical
at finite temperature. This is, however, true only for a long range
correlation with respect to the block structure of the system;
inside one block short range correlations are non-uniform.


\section{Examples}
In this section we are going to illustrate the technique of
calculation of correlation functions for HIC by several examples. As
the technique, despite its clearness and simplicity, leads to rather
cumbersome  expressions, we will present explicit calculations for
the simplest models which have been investigated earlier in a
context of magnetic and thermodynamic
properties\cite{str03,str04,str05,can06,val08,per08,per09,str09,ant09,oha09,
oha10,str10,bel10,roj11a,roj11b,lis11a,lis11b,str11,oha12,cha12,roj12a,ana12,ver12,str12,gal,roj12b}.
One can distinguish at least two classes of HIC: the models where
each block has a left-right symmetry with respect to the interaction
with $\sigma$
spins\cite{str03,str04,str05,can06,val08,per08,per09,str09,ant09,oha10,roj11a},
which implies that the term describing the interaction between block
spins and their left and right $\sigma$-neighbors has the form
$K\left(\sigma_j+\sigma_{j+1}\right)\sum_a S_{j,a}^z$, which is
completely symmetric with respect to the permutation of the
operators $S_{j,a}^z$, and the models which are non symmetric with
respect to left and right $\sigma$-spins on each
block\cite{oha09,bel10,oha12}. It is worth mentioning, that the
experimentally obtained single chain magnet compound with Ising and
Heisenberg bonds, which is an existing example of HIC, belongs to
the second type\cite{chem,Dy_10}. Magnetic properties of the
asymmetric model are vast and complicated, they can exhibit a large
number of ground states with a breaking of translational symmetry or
doubling of the unit cell (block)\cite{oha09, bel10, oha12}. In the
most symmetric case the block Hamiltonian is completely symmetric
with respect to the permutation of all spins in it. Such models are
the simplest ones. For instance, the various variants of the
Heisenberg--Ising diamond chain\cite{can06,str09,roj11a} and
Heisenebrg-Ising tetrahedral\cite{val08} and double-tetrahedral
chains\cite{ant09,oha10} belong to this class of HIC. It is obvious,
that in this case the matrices $\mathbf{\Sigma_a^{\alpha}}$ are the
same for all spins inside one block, or, in other words, in this
case one does not need the additional index $a$.
\begin{figure}
\begin{center}
\includegraphics[width=0.86\columnwidth]{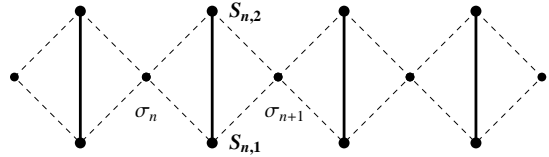}
\caption{\label{fig1} Heisenberg-Ising diamond chain. Solid (dashed)
lines denote quantum XXZ (classical Ising) interaction bonds. All
dashed bonds correspond to the coupling constant $K$, the
interaction between quantum spins is characterized by the coupling
constant $J$ and anisotropy $\Delta$.}
\end{center}
\end{figure}

\subsection{Symmetric diamond-chain}

The symmetric Heisenberg-Ising diamond chain is the simplest example
of HIC (See Figure \ref{fig1}). Its block Hamiltonian has the form:
\begin{eqnarray}\label{ham_dc}
&&\mathcal{H}_n^{DC}=J
\left(\Delta\left(S_{n,1}^xS_{n,2}^x+S_{n,1}^yS_{n,2}^y\right)+S_{n,1}^zS_{n,2}^z
 \right)\nonumber\\
&&-\left(H-K\left( \sigma_n+\sigma_{n+1}\right) \right)\left(
S_{n,1}^z+S_{n,2}^z\right)
\end{eqnarray}
The corresponding transfer-matrix is
\begin{widetext}
\begin{eqnarray}\label{Tdc}
\mathbf{T}^{DC}=2e^{-\beta \frac{J}{4}} \left(
\begin{array}{lcr}
       e^{\beta\frac{H}{2}}\left(\cosh\left[\beta\left(H-K \right)\right]+Q^{DC}\right)  & \cosh\left(\beta H \right)+Q^{DC} \\
       \cosh\left(\beta H \right)+Q^{DC} &   e^{-\beta\frac{H}{2}}\left(\cosh\left[\beta\left(H+K
       \right)\right]+Q^{DC}\right)
      \end{array}\right),
\end{eqnarray}
\end{widetext}
where
\begin{eqnarray}\label{Q_dc}
Q^{DC}=e^{\beta \frac{J}{2}}\cosh\left(
\beta\frac{J\Delta}{2}\right).
\end{eqnarray}
 As the block Hamiltonian is SO(2)-invariant with respect to the
$z$-axis the corresponding $\mathbf{\Sigma}_a^{\alpha}$-matrix is
non-zero only for $\alpha=z$ and, as discussed above, in virtue of
the symmetry with respect to permutation of $\mathbf{S}$-operators
for one block
$\mathbf{\Sigma}_1^{z}=\mathbf{\Sigma}_2^{z}\equiv\mathbf{\Sigma}^{z}$.
The matrix $\mathbf{\Sigma}^{z}$ has a rather simple form:
\begin{eqnarray}\label{Sigma_dc}
&&\mathbf{\Sigma}^{z}=\\
&&e^{-\beta \frac{J}{4}}\left(
\begin{array}{lcr}
e^{\beta\frac{H}{2}}\sinh\left[\beta\left(H-K\right)\right] & \sinh\left(\beta H\right) \\
\sinh\left(\beta H\right) &
e^{-\beta\frac{H}{2}}\sinh\left[\beta\left( H+K\right)\right]
\end{array}\right)\nonumber
\end{eqnarray}
According to Eqs.(\ref{tild_P_ent})-(\ref{corr_Psig}), the non-zero
two-site correlation functions for this model are
\begin{widetext}
\begin{eqnarray}\label{SS_dc}
&&\langle S_{j,a}^z S_{j+r,b}^z\rangle-\langle
S_{j,a}^z\rangle^2=\frac{\left(2\sinh\left(\beta
H\right)M_{\sigma}+\left[e^{\beta
K}\cosh\left(\beta\frac{H}{2}\right)-e^{-\beta
K}\cosh\left(\beta\frac{3H}{2}\right)\right]\sqrt{1-M_{\sigma}^2}
\right)^2}{16 \left(\sinh^2\left( \beta K\right)+2 Q^{DC} \left(
\cosh\left(\beta K\right)-1\right)\cosh\left(\beta H\right)
\right)}e^{-\frac{r}{\xi}},\\
&&\langle S_{j,a}^z \sigma_{j+r}^z\rangle-\langle
S_{j,a}^z\rangle\langle
\sigma_j\rangle=-\frac{e^{-\beta\frac{J}{4}}\left\{2\sinh\left(\beta
H\right)M_{\sigma}\sqrt{1-M_{\sigma}^2}+\left[e^{\beta
K}\cosh\left(\beta\frac{H}{2}\right)-e^{-\beta
K}\cosh\left(\beta\frac{3H}{2}\right)\right]\left(1-M_{\sigma}^2\right)^2
\right\}}{4\lambda_1^{DC}}e^{-\frac{r}{\xi}} \nonumber
\end{eqnarray}
\end{widetext}
where
\begin{eqnarray}\label{Ms}
\langle
S_{j,a}^z\rangle=\frac{1}{\mathcal{Z}}{{\mathbf{Tr}}}\left({\mathbf{\Lambda}}^{N-1}{\tilde{\mathbf{\Sigma^z}}}\right)=\frac{1}{2}M_S
\end{eqnarray}
is the partial magnetization of the sublattice of quantum spins.
$M_{\sigma}$ is the magnetization of $\sigma$-sublattice obtained
according to Eq.(\ref{M}) from transfer-matrix (\ref{Tdc}) and
$\lambda_1^{DC}$ is the largest eigenvalue of the transfer-matrix
(\ref{Tdc}). The eigenvalues have the following form:
\begin{widetext}
\begin{eqnarray}\label{l1_dc}
\lambda_{1,2}^{DC}=e^{-\beta\frac{J}{4}}&&\left\{ e^{-\beta
K}\cosh\left(\beta\frac{3 H}{2} \right)+\left(2 Q^{DC}+e^{\beta
K}\right)\cosh\left( \beta\frac{H}{2}\right)\right. \\
&&\pm\left.\sqrt{\left[e^{-\beta K}\sinh\left(\beta\frac{3 H}{2}
\right)+\left(2 Q^{DC}-e^{\beta K}\right)\sinh\left(
\beta\frac{H}{2}\right) \right]^2+4\left( \cosh\left(\beta
H\right)+Q^{DC}\right)^2}\right\}.\nonumber
\end{eqnarray}
\end{widetext}
\subsection{Asymmetric sawtooth chain}
\begin{figure}
\begin{center}
\includegraphics[width=0.96\columnwidth]{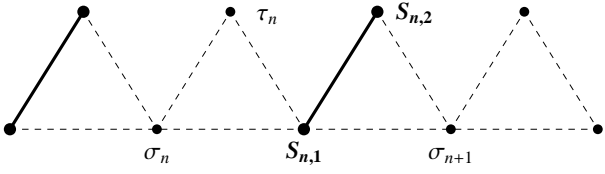}
\caption{\label{fig2} Heisenberg-Ising sawtooth chain with a pair of
quantum spins in one block. The unit cell consists of 4 spins,
$\mathbf{S}_{j,1}, \mathbf{S}_{j,2}, \sigma_j$ and $\tau_j$. All
bonds denoted by dashed lines are of Ising type and the
corresponding coupling constant is $K$, while the $XXZ$-interaction
corresponding to the quantum bond (solid line) is given by the
coupling and anisotropy $J$ and $\Delta$.}
\end{center}
\end{figure}
The next example we are going to consider is the Heisenberg-Ising
sawtooth chain introduced in Refs. \onlinecite{oha09} and
\onlinecite{bel10}. The lattice is depicted in Figure \ref{fig2}.
This model is quite convenient for illustrating the asymmetric role
of quantum spins from one block as well as for the demonstration of
the role of the additional classical spin included in the block.
Here the block Hamiltonian contains two parts, one with quantum
operators and another one - without them. Thus, one can write down
the block Hamiltonian for the sawtooth HIC with 2 quantum spins in
one block in the following way:
\begin{eqnarray}\label{ham_sc}
&&\mathcal{H}_n^{SC}=\mathcal{H}^q_n+K\sigma_n\tau_n-H\tau_n,\\
&&\mathcal{H}^q_n=J
\left(\Delta\left(S_{n,1}^xS_{n,2}^x+S_{n,1}^yS_{n,2}^y\right)+S_{n,1}^zS_{n,2}^z
 \right)\nonumber\\
 &&-\left(H-K\left( \sigma_n+\sigma_{n+1}+\tau_n\right)
 \right)S_{n,1}^z-\left(H-K\sigma_{n+1} \right)S_{n,2}^z.\nonumber
\end{eqnarray}
Here all notations are explained in Figure \ref{fig2}. In contrast
to the case considered previously in Refs. \onlinecite{oha09} and
\onlinecite{bel10}, here, for the sake of simplicity, we consider
all Ising bonds with the same coupling constant $K$. As the block
Hamiltonian contains also the Ising spin $\tau$, in order to obtain
the entries of transfer matrix one needs, after taking the trace
over quantum states of the pair of $\mathbf{S}$-spin, to sum over
the two possible values of the classical spin $\tau$. The entries of
the resulting classical transfer matrix are:
\begin{widetext}
\begin{eqnarray}\label{T_sc}
&&T_{++}^{SC}=2e^{-\beta\frac{J}{4}}\left\{e^{\beta\left(H-\frac{K}{4}\right)}\left[\cosh\left(\beta\left(H-K\right)\right)
+Q_2^{SC}\right]+e^{\beta\frac{K}{4}}\left[\cosh\left(\beta\left(H-\frac{K}{2}\right)\right)
+Q_1^{SC}\right]\right\},\\
&&T_{+-}^{SC}=2e^{-\beta\frac{J}{4}}\left\{e^{\frac{\beta}{2}\left(H-\frac{K}{2}\right)}\left[\cosh\left(\beta
H\right)+Q_2^{SC}\right]+e^{-\frac{\beta}{2}\left(H-\frac{K}{2}\right)}\left[\cosh\left(\beta\left(H+\frac{K}{2}\right)\right)
+Q_1^{SC}\right]\right\},\nonumber \\
&&T_{-+}^{SC}=2e^{-\frac{\beta}{4}}\left\{e^{-\frac{\beta}{2}\left(H+\frac{K}{2}\right)}\left[\cosh\left(\beta
H\right)
+Q_2^{SC}\right]+e^{\frac{\beta}{2}\left(H+\frac{K}{2}\right)}\left[\cosh\left(\beta\left(
H-\frac{K}{2}\right)\right)
+Q_1^{SC}\right]\right\},\nonumber \\
&&T_{--}^{SC}=2e^{-\beta\frac{J}{4}}\left\{e^{-\beta\left(H+\frac{K}{4}\right)}\left[\cosh\left(\beta\left(H+K\right)\right)
+Q_2^{SC}\right]+e^{\beta\frac{K}{4}}\left[\cosh\left(\beta\left(H+\frac{K}{2}\right)\right)
+Q_1^{SC}\right]\right\}, \nonumber
\end{eqnarray}
\end{widetext}
where
\begin{eqnarray}\label{Qsc}
&&Q_1^{SC}=e^{\beta\frac{J}{2}}\cosh\left(\beta\frac{J\Delta}{2}\right),\\
&&Q_2^{SC}=e^{\beta\frac{J}{2}}\cosh\left(\beta\frac{\sqrt{K^2+J^2\Delta^2}}{2}\right).\nonumber
\end{eqnarray}
The block Hamiltonian (\ref{ham_sc}) is not symmetric with respect
the to permutation of the quantum spin operators
${\mathbf{S}}_{n,1}$ and ${\mathbf{S}}_{n,2}$, thus, there are two
matrices representing the $z$-component of the corresponding
spin-operators using for the calculation of the correlation
functions. In virtue of the SO(2)-invariance of Eq.(\ref{ham_sc})
the corresponding matrices for other spin components are identically
equal to zero. Let us write down the entries of the
${\mathbf{\Sigma}}_1^z$ and ${\mathbf{\Sigma}}_2^z$ matrices for the
Heisenberg-Ising sawtooth chain.
\begin{widetext}
\begin{eqnarray}\label{Sigma1}
&&\left({\mathbf{\Sigma}_{1,2}^z}\right)_{++}=e^{-\beta\frac{J}{4}}\left\{
       e^{\beta\frac{K}{4}}\cosh\left( \beta\left(H-\frac{K}{2}\right)\right)
       +e^{\beta\left(H-\frac{K}{4}\right)}\left(\cosh\left(\beta\left(H-K\right)\right)\mp W\right)\right\},\\
&&\left({\mathbf{\Sigma}_{1,2}^z}\right)_{+-}=
e^{-\beta\frac{J}{4}}\left\{e^{-\frac{\beta}{2}\left(H-\frac{K}{2}\right)}
       \cosh\left( \beta\left(H+\frac{K}{2}\right)\right)
       +e^{\frac{\beta}{2}\left(H-\frac{K}{2}\right)}\left(\cosh\left(\beta H\right)\mp W\right)\right\},\nonumber \\
&&\left({\mathbf{\Sigma}_{1,2}^z}\right)_{-+}=
e^{-\beta\frac{J}{4}}\left\{e^{\frac{\beta}{2}\left(H+\frac{K}{2}\right)}
       \cosh\left( \beta\left(H-\frac{K}{2}\right)\right)
       +e^{-\frac{\beta}{2}\left(H+\frac{K}{2}\right)}\left(\cosh\left(\beta H\right)\pm W\right)\right\},\nonumber \\
&&\left({\mathbf{\Sigma}_{1,2}^z}\right)_{--}=e^{-\beta\frac{J}{4}}\left\{
       e^{\beta\frac{K}{4}}\cosh\left( \beta\left(H+\frac{K}{2}\right)\right)
       +e^{-\beta\left(H+\frac{K}{4}\right)}\left(\cosh\left(\beta\left(H+K\right)\right)\pm W\right)\right\},
       \nonumber
\end{eqnarray}
\end{widetext}
where
\begin{eqnarray}\label{W}
W=K\frac{e^{\beta\frac{J}{2}}\sinh\left(\beta\frac{\sqrt{K^2+J^2\Delta^2}}{2}\right)}{\sqrt{K^2+J^2\Delta^2}}.
\end{eqnarray}
and upper(lower) signs correspond to
${\mathbf{\Sigma}}_1^z$(${\mathbf{\Sigma}}_2^z$). The matrix
$\mathbf{\tau}$ in terms of which the expectation values and
correlation function of the classical spin from block Hamiltonian
are expressed has the following entries:
\begin{widetext}
\begin{eqnarray}\label{tau}
&&\tau_{++}=e^{-\beta\frac{J}{4}}\left\{e^{\beta\left(H-\frac{K}{4}\right)}\left[\cosh\left(\beta\left(H-K\right)\right)
+Q_2^{SC}\right]-e^{\beta\frac{K}{4}}\left[\cosh\left(\beta\left(H-\frac{K}{2}\right)\right)
+Q_1^{SC}\right]\right\},\\
&&\tau_{+-}=e^{-\beta\frac{J}{4}}\left\{e^{\frac{\beta}{2}\left(H-\frac{K}{2}\right)}\left[\cosh\left(\beta
H\right)+Q_2^{SC}\right]-e^{-\frac{\beta}{2}\left(H-\frac{K}{2}\right)}\left[\cosh\left(\beta\left(H+\frac{K}{2}\right)\right)
+Q_1^{SC}\right]\right\},\nonumber \\
&&\tau_{-+}=e^{-\frac{\beta}{4}}\left\{-e^{-\frac{\beta}{2}\left(H+\frac{K}{2}\right)}\left[\cosh\left(\beta
H\right)
+Q_2^{SC}\right]+e^{\frac{\beta}{2}\left(H+\frac{K}{2}\right)}\left[\cosh\left(\beta\left(
H-\frac{K}{2}\right)\right)
+Q_1^{SC}\right]\right\},\nonumber \\
&&\tau_{--}=e^{-\beta\frac{J}{4}}\left\{-e^{-\beta\left(H+\frac{K}{4}\right)}\left[\cosh\left(\beta\left(H+K\right)\right)
+Q_2^{SC}\right]+e^{\beta\frac{K}{4}}\left[\cosh\left(\beta\left(H+\frac{K}{2}\right)\right)
+Q_1^{SC}\right]\right\}. \nonumber
\end{eqnarray}
\end{widetext}
All possible two-spin correlation functions (except the
$\sigma\sigma$, which is uniform for all HIC and is given by Eq.
(\ref{cor_sigma_3_xi})) of the system are listed below,
\begin{widetext}
\begin{eqnarray}\label{SC_all_corr}
&&\langle S_{j,1}^z S_{j+r,1}^z\rangle-\langle
S_{j,1}^z\rangle^2=\frac{\left(\sum_{k=1}^4\left[A_k^1\sinh\left(\beta\frac{k
H}{2}\right)+B_k^1\cosh\left(\beta\frac{k
H}{2}\right)\right]\right)\left(\sum_{k=1}^4\left[A_k^3\sinh\left(\beta\frac{k
H}{2}\right)+B_k^3\cosh\left(\beta\frac{k
H}{2}\right)\right]\right)}{16\sum_{k=0}^2 D_k\cosh
\left(\beta k H\right)}e^{-\frac{r}{\xi}},\\
&&\langle S_{j,2}^z S_{j+r,2}^z\rangle-\langle
S_{j,2}^z\rangle^2=\frac{\left(\sum_{k=1}^4\left[A_k^2\sinh\left(\beta\frac{k
H}{2}\right)+B_k^2\cosh\left(\beta\frac{k
H}{2}\right)\right]\right)\left(\sum_{k=1}^4\left[A_k^4\sinh\left(\beta\frac{k
H}{2}\right)+B_k^4\cosh\left(\beta\frac{k
H}{2}\right)\right]\right)}{16\sum_{k=0}^2 D_k\cosh
\left(\beta k H\right)}e^{-\frac{r}{\xi}}, \nonumber\\
&&\langle S_{j,1}^z S_{j+r,2}^z\rangle-\langle
S_{j,1}^z\rangle\langle
S_{j,2}^z\rangle=\frac{\left(\sum_{k=1}^4\left[A_k^1\sinh\left(\beta\frac{k
H}{2}\right)+B_k^1\cosh\left(\beta\frac{k
H}{2}\right)\right]\right)\left(\sum_{k=1}^4\left[A_k^4\sinh\left(\beta\frac{k
H}{2}\right)+B_k^4\cosh\left(\beta\frac{k
H}{2}\right)\right]\right)}{16\sum_{k=0}^2 D_k\cosh
\left(\beta k H\right)}e^{-\frac{r}{\xi}}, \nonumber\\
&&\langle S_{j,2}^z S_{j+r,1}^z\rangle-\langle S_{j,1}^z\rangle
\langle
S_{j,2}^z\rangle=\frac{\left(\sum_{k=1}^4\left[A_k^2\sinh\left(\beta\frac{k
H}{2}\right)+B_k^2\cosh\left(\beta\frac{k
H}{2}\right)\right]\right)\left(\sum_{k=1}^4\left[A_k^3\sinh\left(\beta\frac{k
H}{2}\right)+B_k^3\cosh\left(\beta\frac{k
H}{2}\right)\right]\right)}{16\sum_{k=0}^2 D_k\cosh
\left(\beta k H\right)}e^{-\frac{r}{\xi}},\nonumber \\
&&\langle S_{j,1}^z \sigma_{j+r}\rangle-\langle
S_{j,1}^z\rangle\langle
\sigma_j\rangle=-\frac{e^{-\beta\frac{J}{4}}\sqrt{1-M_{\sigma}^2}\sum_{k=1}^4\left(A_k^1\sinh\left(\beta\frac{k
H}{2}\right)+B_k^1\cosh\left(\beta\frac{k
H}{2}\right)\right)}{4\lambda_1^{SC}}e^{-\frac{r}{\xi}}, \nonumber\\
&&\langle S_{j,2}^z \sigma_{j+r}\rangle-\langle
S_{j,2}^z\rangle\langle
\sigma_j\rangle=-\frac{e^{-\beta\frac{J}{4}}\sqrt{1-M_{\sigma}^2}\sum_{k=1}^4\left(A_k^2\sinh\left(\beta\frac{k
H}{2}\right)+B_k^2\cosh\left(\beta\frac{k
H}{2}\right)\right)}{4\lambda_1^{SC}}e^{-\frac{r}{\xi}}, \nonumber\\
&&\langle S_{j,1}^z \tau_{j+r}\rangle-\langle
S_{j,1}^z\rangle\langle
\tau_j\rangle=\frac{\left(\sum_{k=1}^4\left[A_k^1\sinh\left(\beta\frac{k
H}{2}\right)+B_k^1\cosh\left(\beta\frac{k
H}{2}\right)\right]\right)\left(\sum_{k=0}^4\left[A_k^6\sinh\left(\beta\frac{k
H}{2}\right)+B_k^6\cosh\left(\beta\frac{k
H}{2}\right)\right]\right)}{16\sum_{k=0}^2 D_k\cosh
\left(\beta k H\right)}e^{-\frac{r}{\xi}}, \nonumber\\
&&\langle S_{j,2}^z \tau_{j+r}\rangle-\langle
S_{j,2}^z\rangle\langle
\tau_j\rangle=\frac{\left(\sum_{k=1}^4\left[A_k^2\sinh\left(\beta\frac{k
H}{2}\right)+B_k^2\cosh\left(\beta\frac{k
H}{2}\right)\right]\right)\left(\sum_{k=0}^4\left[A_k^6\sinh\left(\beta\frac{k
H}{2}\right)+B_k^6\cosh\left(\beta\frac{k
H}{2}\right)\right]\right)}{16\sum_{k=0}^2 D_k\cosh
\left(\beta k H\right)}e^{-\frac{r}{\xi}}, \nonumber\\
&&\langle \sigma_{j}S_{j+r,1}^z\rangle-\langle
S_{j,1}^z\rangle\langle
\sigma_j\rangle=-\frac{e^{-\beta\frac{J}{4}}\sqrt{1-M_{\sigma}^2}\sum_{k=1}^4\left(A_k^3\sinh\left(\beta\frac{k
H}{2}\right)+B_k^3\cosh\left(\beta\frac{k
H}{2}\right)\right)}{4\lambda_1^{SC}}e^{-\frac{r}{\xi}}, \nonumber\\
&&\langle \sigma_{j}S_{j+r,2}^z\rangle-\langle
S_{j,2}^z\rangle\langle
\sigma_j\rangle=-\frac{e^{-\beta\frac{J}{4}}\sqrt{1-M_{\sigma}^2}\sum_{k=1}^4\left(A_k^4\sinh\left(\beta\frac{k
H}{2}\right)+B_k^4\cosh\left(\beta\frac{k
H}{2}\right)\right)}{4\lambda_1^{SC}}e^{-\frac{r}{\xi}}, \nonumber\\
&&\langle \sigma_j \tau_{j+r}\rangle-\langle \sigma_j\rangle\langle
\tau_j\rangle=-\frac{e^{-\beta\frac{J}{4}}\sqrt{1-M_{\sigma}^2}\sum_{k=0}^4\left[A_k^6\sinh\left(\beta\frac{k
H}{2}\right)+B_k^6\cosh\left(\beta\frac{k
H}{2}\right)\right]}{4\lambda_1^{SC}}e^{-\frac{r}{\xi}}, \nonumber\\
&&\langle  \tau_{j} S_{j+r,1}^z\rangle-\langle
S_{j,1}^z\rangle\langle
\tau_j\rangle=\frac{\left(\sum_{k=0}^4\left[A_k^5\sinh\left(\beta\frac{k
H}{2}\right)+B_k^5\cosh\left(\beta\frac{k
H}{2}\right)\right]\right)\left(\sum_{k=1}^4\left[A_k^3\sinh\left(\beta\frac{k
H}{2}\right)+B_k^3\cosh\left(\beta\frac{k
H}{2}\right)\right]\right)}{16\sum_{k=0}^2 D_k\cosh
\left(\beta k H\right)}e^{-\frac{r}{\xi}}, \nonumber\\
&&\langle  \tau_{j} S_{j+r,2}^z\rangle-\langle
S_{j,2}^z\rangle\langle
\tau_j\rangle=\frac{\left(\sum_{k=0}^4\left[A_k^5\sinh\left(\beta\frac{k
H}{2}\right)+B_k^5\cosh\left(\beta\frac{k
H}{2}\right)\right]\right)\left(\sum_{k=1}^4\left[A_k^4\sinh\left(\beta\frac{k
H}{2}\right)+B_k^4\cosh\left(\beta\frac{k
H}{2}\right)\right]\right)}{16\sum_{k=0}^2 D_k\cosh
\left(\beta k H\right)}e^{-\frac{r}{\xi}}, \nonumber\\
&&\langle \tau_{j} \sigma_{j+r}\rangle-\langle
\sigma_j\rangle\langle
\tau_j\rangle=-\frac{e^{-\beta\frac{J}{4}}\sqrt{1-M_{\sigma}^2}\sum_{k=0}^4\left[A_k^5\sinh\left(\beta\frac{k
H}{2}\right)+B_k^5\cosh\left(\beta\frac{k
H}{2}\right)\right]}{4\lambda_1^{SC}}e^{-\frac{r}{\xi}}, \nonumber\\
&&\langle \tau_j \tau_{j+r}\rangle-\langle
\tau_j\rangle^2=\frac{\left(\sum_{k=0}^4\left[A_k^5\sinh\left(\beta\frac{k
H}{2}\right)+B_k^5\cosh\left(\beta\frac{k
H}{2}\right)\right]\right)\left(\sum_{k=0}^4\left[A_k^6\sinh\left(\beta\frac{k
H}{2}\right)+B_k^6\cosh\left(\beta\frac{k
H}{2}\right)\right]\right)}{16\sum_{k=0}^2 D_k\cosh \left(\beta k
H\right)}e^{-\frac{r}{\xi}}, \nonumber
\end{eqnarray}
\end{widetext}
where the eigenvalues of the transfer-matrix (\ref{T_sc}) are given
by the following expressions:
\begin{widetext}
\begin{eqnarray}\label{lambda_sc}
\lambda_{1,2}^{SC}=e^{-\beta\frac{J}{4}}\left(\sum_{k=0}^2 S_k
\cosh\left(\beta k H\right)\pm\sqrt{\left(\sum_{k=0}^2 S_k
\cosh\left(\beta k H\right)\right)^2-4\sum_{k=0}^2 D_k
\cosh\left(\beta k H\right)} \right).
\end{eqnarray}
\end{widetext}
 The coefficients in Eqs.(\ref{SC_all_corr}) and (\ref{lambda_sc})
 are
\begin{widetext}
\begin{eqnarray}\label{coef}
&&A_1^1=e^{\beta\frac{3 K}{4}}-e^{-\beta\frac{K}{4}}\left(1+2 W
M_{\sigma}\right),\;\;
A_2^1=2e^{\beta\frac{K}{4}}\sinh\left(\beta\frac{K}{2}\right)\sqrt{1-M_{\sigma}^2},\;\;
A_3^1=0,\;\; A_4^1=-e^{-\beta\frac{5 K}{4}}\sqrt{1-M_{\sigma}^2},\\
&&B_1^1=\left(e^{\beta\frac{3 K}{4}}+e^{-\beta\frac{
K}{4}}\right)M_{\sigma}-2 e^{-\beta\frac{K}{4}}W,\;\; B_2^1=2
We^{-\beta\frac{K}{4}} \sqrt{1-M_{\sigma}^2},\;\;
B_3^1=2e^{-\beta\frac{K}{4}}M_{\sigma},\;\; B_4^1=0,\nonumber \\
&& A_1^2=e^{\beta\frac{3 K}{4}}-e^{-\beta\frac{K}{4}}\left(1-2 W
M_{\sigma}\right),\;\; A_2^2=A_2^1,\;\;
A_3^2=0,\;\; A_4^2=A_4^1,\nonumber \\
&&B_1^2=\left(e^{\beta\frac{3 K}{4}}+e^{-\beta\frac{
K}{4}}\right)M_{\sigma}+2 e^{-\beta\frac{K}{4}}W,\;\; \;B_2^2=-2
We^{-\beta\frac{K}{4}} \sqrt{1-M_{\sigma}^2},\;\; B_3^2=B_3^1,\;\;
B_4^2=0,\nonumber\\
&&A_1^3=-e^{\beta\frac{3 K}{4}}+e^{-\beta\frac{K}{4}}\left(1-2 W
M_{\sigma}\right),\;\; A_2^3=A_2^1,\;\;A_3^3=0,\;\;
A_4^3=A_4^1,\nonumber\\
&&B_1^3=B_1^2,\;\; B_2^3=B_2^1,\;\; B_3^3=B_2^1,\;\;
B_4^3=0,\nonumber\\
&& A_1^4=-e^{\beta\frac{3 K}{4}}+e^{-\beta\frac{K}{4}}\left(1+2 W
M_{\sigma}\right),\;\; A_2^4=A_2^1,\;\; A_3^4=0,\;\; A_4^4=A_4^1\nonumber \\
&&B_1^4=B_1^1,\;\; B_2^4=-B_1^2,\;\; B_3^4=B_3^1,\;\;
B_4^4=0,\nonumber\\
&&A_1^5=\left(2e^{-\beta\frac{K}{4}}Q_2^{SC}+2e^{\beta\frac{K}{4}}Q_1^{SC}-e^{-\beta\frac{K}{4}}-e^{\beta\frac{3K}{4}}\right)M_{\sigma},\;\;
A_2^5=0,\;\; A_3^5=2e^{-\beta\frac{K}{4}}M_{\sigma},\;\; A_4^5=0\nonumber \\
&&B_0^5=\left(2
e^{\beta\frac{K}{4}}Q_1^{SC}-e^{\beta\frac{3K}{4}}\right)\sqrt{1-M_{\sigma}^2},\;\;
B_1^5=e^{-\beta\frac{K}{4}}\left(1+2
Q_2^{SC}\right)-2e^{\beta\frac{K}{4}}-e^{\beta\frac{3
K}{4}},\nonumber\\
&&B_2^5=2\left(e^{\beta\frac{K}{4}}\cosh\left(\beta\frac{K}{2}\right)-e^{-\beta\frac{K}{4}}Q_2^{SC}\right)\sqrt{1-M_{\sigma}^2},\;\;
B_3^5=0,\;\; B_4^5=-e^{-\beta\frac{5K}{4}}\sqrt{1-M_{\sigma}^2}\nonumber \\
&&A_1^6=A_1^5,\;\; A_2^6=0,\;\; A_3^6=A_3^5,\;\; A_4^6=0,\nonumber\\
&&B_0^6=B_0^5,\;\; B_1^6=-B_1^5,\;\; B_2^6=B_2^5,\;\; B_3^6=0, B_4^6=B_4^5,\nonumber\\
 &&D_0=e^{-\beta\frac{K}{2}}\sinh^2\left(\beta
K\right)+\left(e^{\beta K}-1\right)Q_1^{SC}-2
Q_2^{SC}\sinh\left(\beta\frac{K}{2}\right),\nonumber \\
&&D_1=2e^{-\beta\frac{K}{2}}\left(\cosh\left(\beta
K\right)-1\right)Q_2^{SC}+\cosh\left(\beta\frac{3
K}{2}\right)-\cosh\left(\beta\frac{K}{2}\right),\;\; D_2=2
Q_2^{SC}\sinh\left(\beta\frac{K}{2}\right)+\left(e^{-\beta
K}-1\right)Q_1^{SC},\nonumber\\
&&S_0=e^{\beta\frac{3K}{4}}+2 e^{\beta\frac{K}{4}}Q_1^{SC},\;\;
S_1=2\left(e^{-\beta\frac{K}{4}}
Q_2^{SC}+e^{\beta\frac{K}{4}}\cosh\left(\beta\frac{K}{2}\right)\right),\;\;
S_2=e^{-\beta\frac{5K}{4}}.\nonumber
\end{eqnarray}
\end{widetext}

\begin{figure*}[tb!]
\begin{tabular}{cc}
\includegraphics[width=\columnwidth]{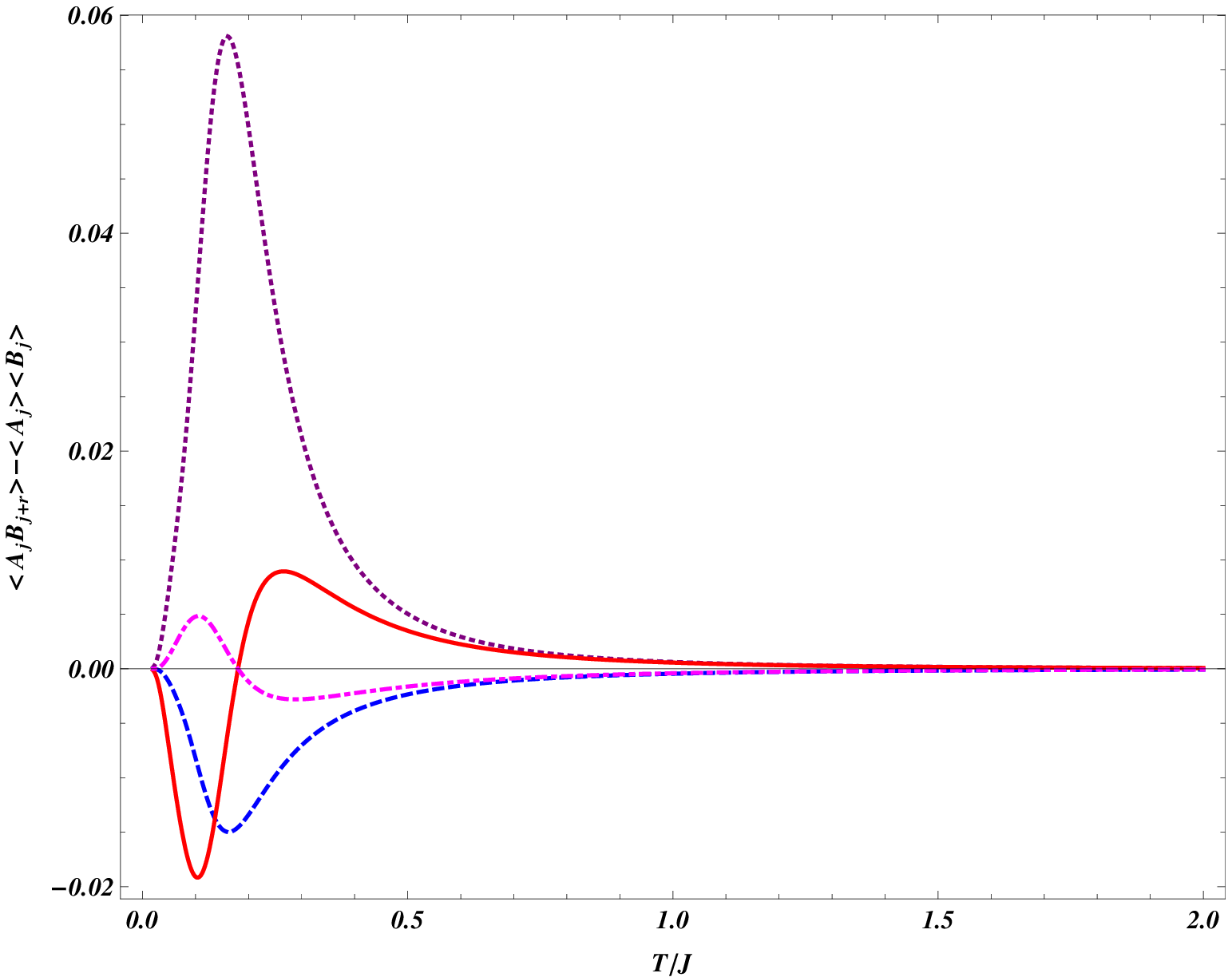}&
\includegraphics[width=\columnwidth]{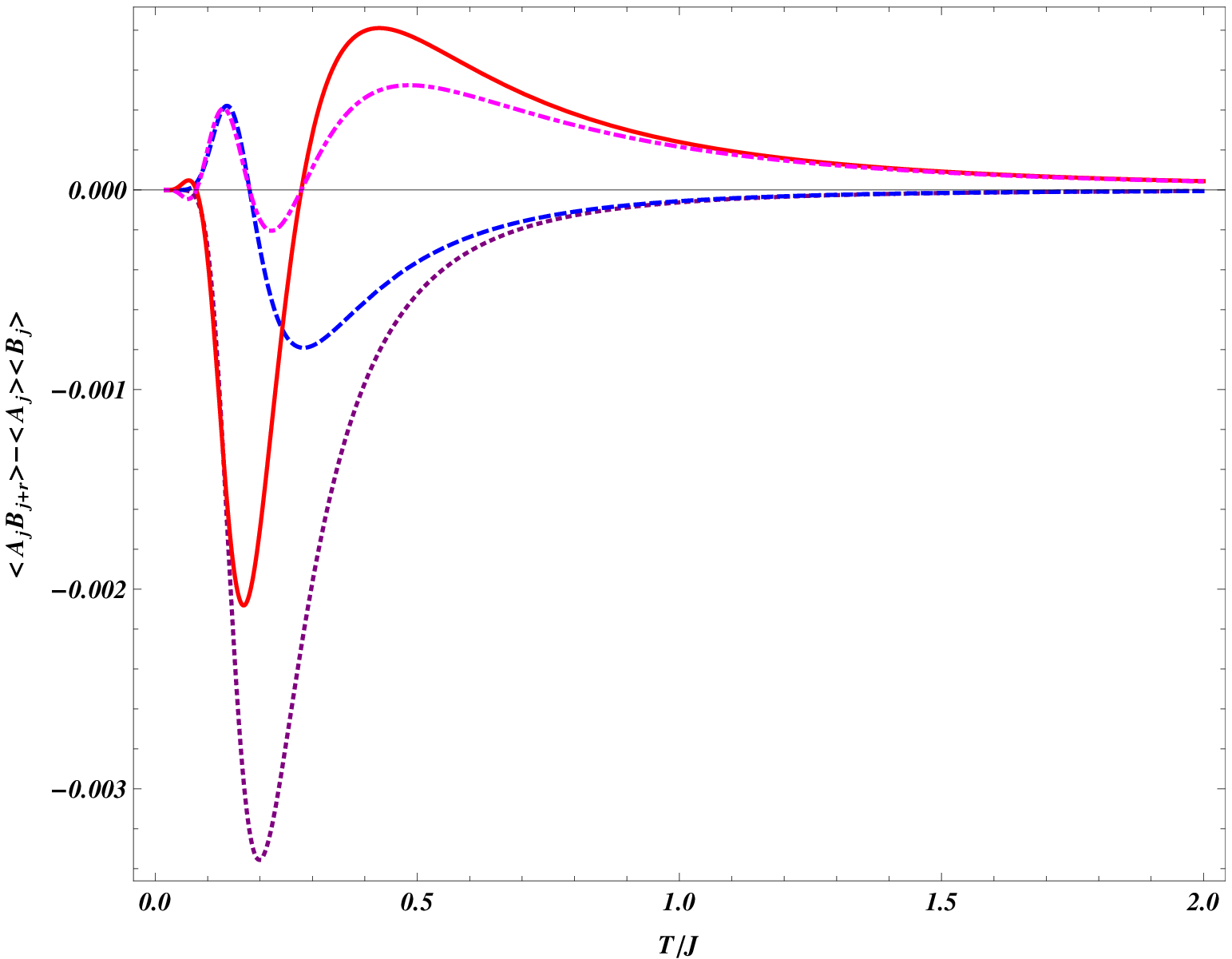}
\end{tabular}
\begin{tabular}{cc}
\includegraphics[width=\columnwidth]{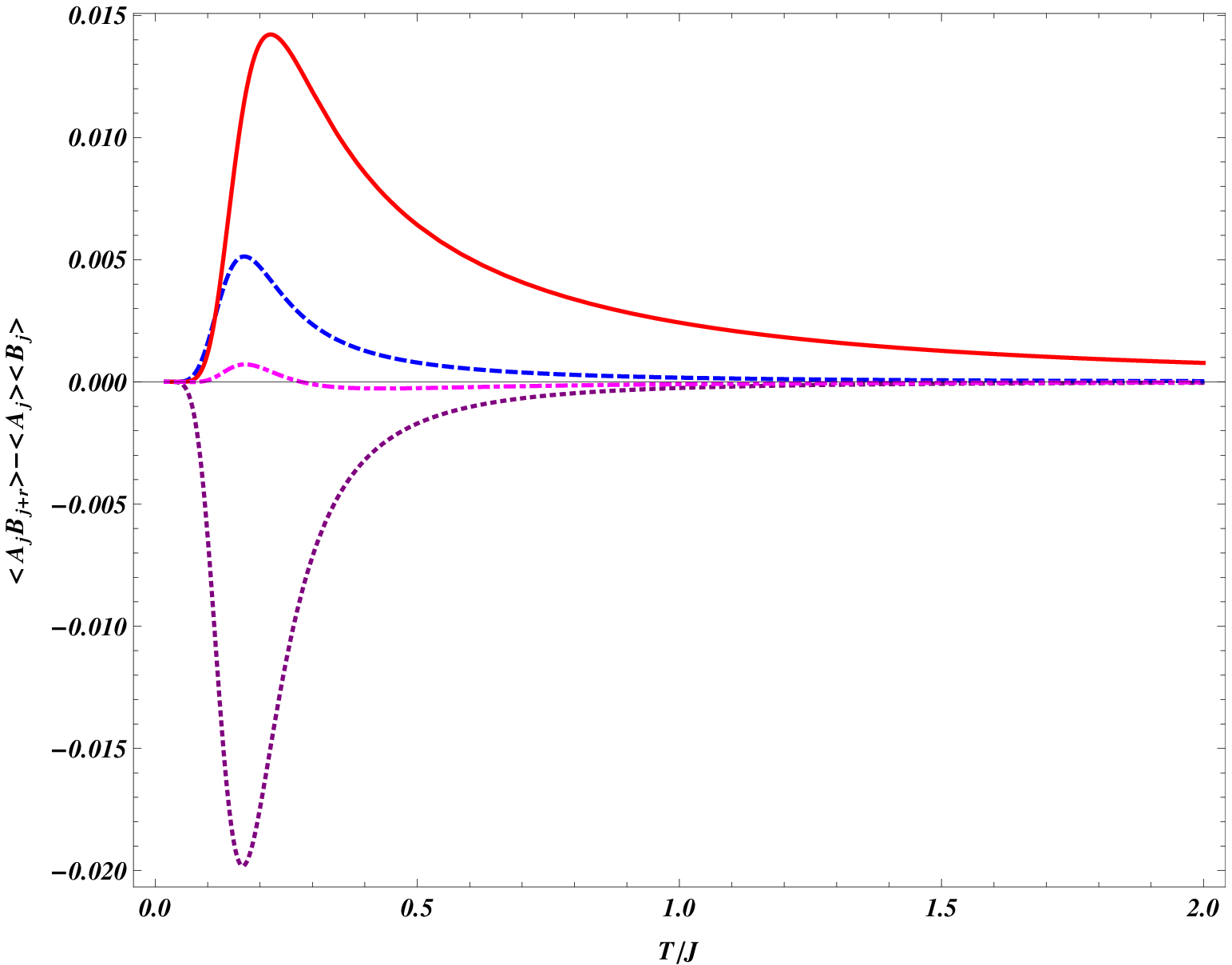}
\includegraphics[width=\columnwidth]{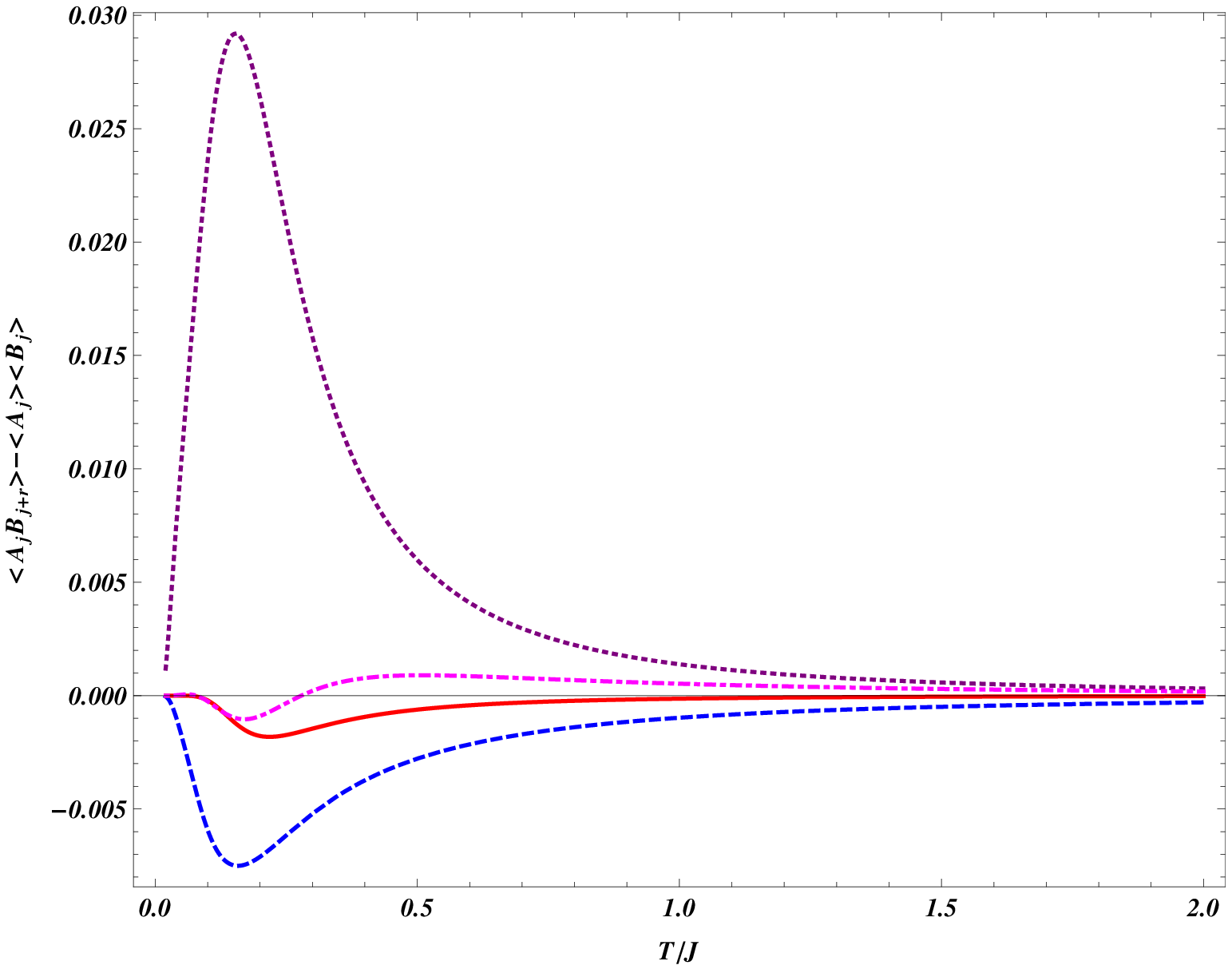}
\end{tabular}
\caption{\label{fig3}(Color online) The temperature dependence of
the correlation functions for the asymmetric sawtooth chain for
$K/J=0.5, H/J=0.5, \Delta=0.3$ at $r=1$ (See Eq. \ref{SC_all_corr}).
Upper left panel: $\langle S_{j,1}^z S_{j+r,1}^z\rangle-\langle
S_{j,1}^z\rangle^2$ - dotted (purple); $\langle S_{j,1}^z
S_{j+r,2}^z\rangle-\langle S_{j,1}^z\rangle\langle S_{j,2}^z\rangle$
- dashed (blue); $\langle S_{j,2}^z S_{j+r,1}^z\rangle-\langle
S_{j,1}^z\rangle\langle S_{j,2}^z\rangle$ - solid (red); $\langle
S_{j,2}^z S_{j+r,2}^z\rangle-\langle S_{j,2}^z\rangle^2$ -
dot-dashed (magenta). Upper right panel: $\langle S_{j,1}^z
\sigma_{j+r}\rangle-\langle S_{j,1}^z\rangle\langle \sigma_j\rangle$
- dotted (purple); $\langle S_{j,2}^z \sigma_{j+r}\rangle-\langle
S_{j,2}^z\rangle\langle \sigma_j\rangle$ - dashed (blue); $\langle
S_{j,1}^z \tau_{j+r}\rangle-\langle S_{j,1}^z\rangle\langle
\tau_j\rangle$ - solid (red); $\langle S_{j,2}^z
\tau_{j+r}\rangle-\langle S_{j,2}^z\rangle\langle \tau_j\rangle $ -
dot dashed (magenta). Lower left panel: $\langle
\sigma_{j}S_{j+r,1}^z\rangle-\langle S_{j,1}^z\rangle\langle
\sigma_j\rangle$ - dotted (purple), $\langle
\sigma_{j}S_{j+r,2}^z\rangle-\langle S_{j,2}^z\rangle\langle
\sigma_j\rangle$ - dashed (blue),
$\langle\sigma_{j}\sigma_{j+r,1}^z\rangle-\langle \sigma_j\rangle^2$
- solid (red); $\langle \sigma_{j}\tau_{j+r,1}^z\rangle-\langle
\sigma_j\rangle\langle \tau_j\rangle$ - dot-dashed (magenta). Lower
right panel: $\langle  \tau_{j} S_{j+r,1}^z\rangle-\langle
S_{j,1}^z\rangle\langle \tau_j\rangle$ - dotted (purple); $\langle
\tau_{j} S_{j+r,2}^z\rangle-\langle S_{j,2}^z\rangle\langle
\tau_j\rangle$ - dashed (blue); $\langle  \tau_{j}
\sigma_{j+r}^z\rangle-\langle \sigma_{j}\rangle\langle
\tau_j\rangle$ - solid (red); $\langle  \tau_{j}
\tau_{j+r}\rangle-\langle \tau_j\rangle^2$ - dot-dashed (magenta).}
\end{figure*}

\begin{figure*}[tbph]
\begin{tabular}{cc}
\includegraphics[width=\columnwidth]{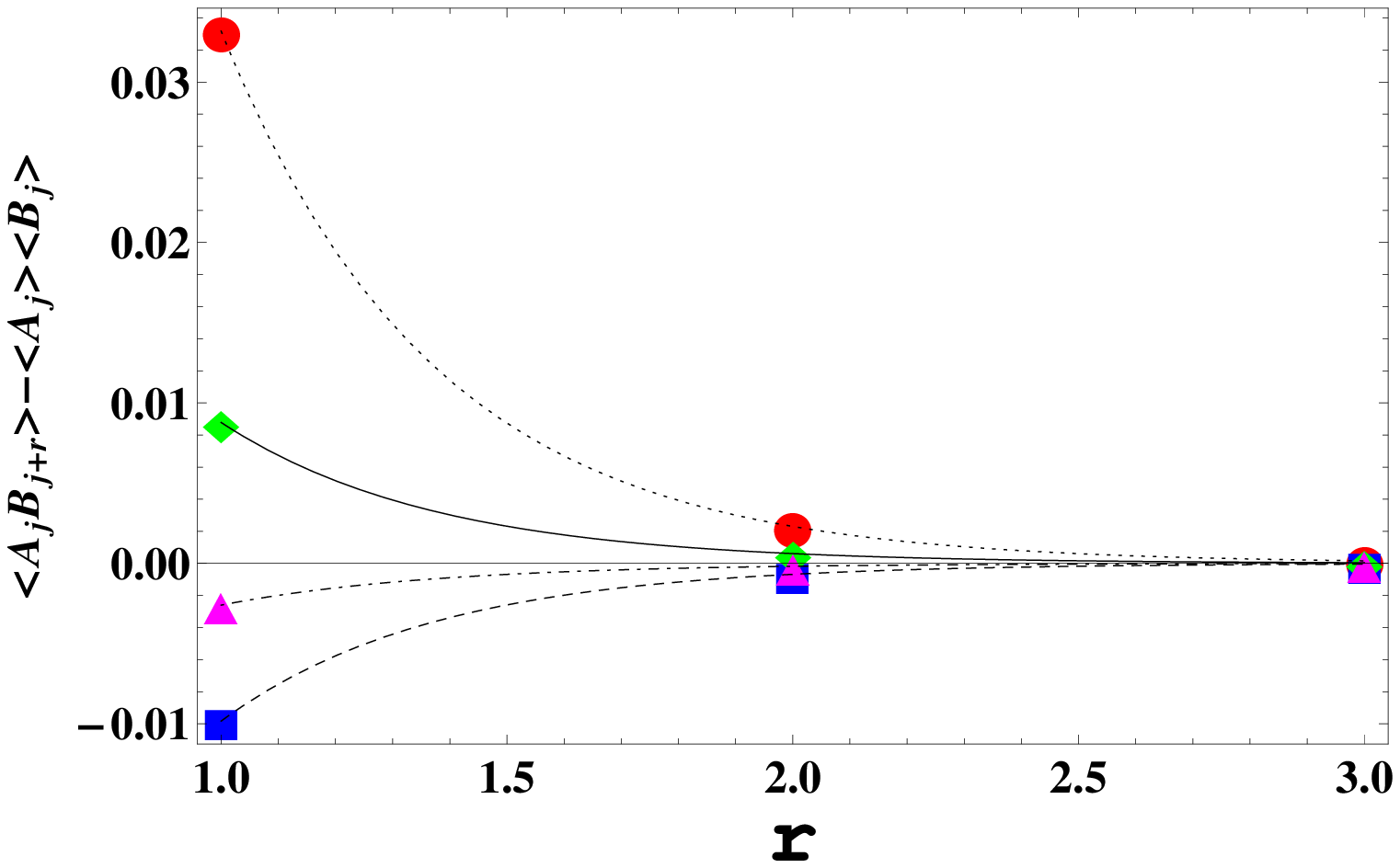}&
\includegraphics[width=\columnwidth]{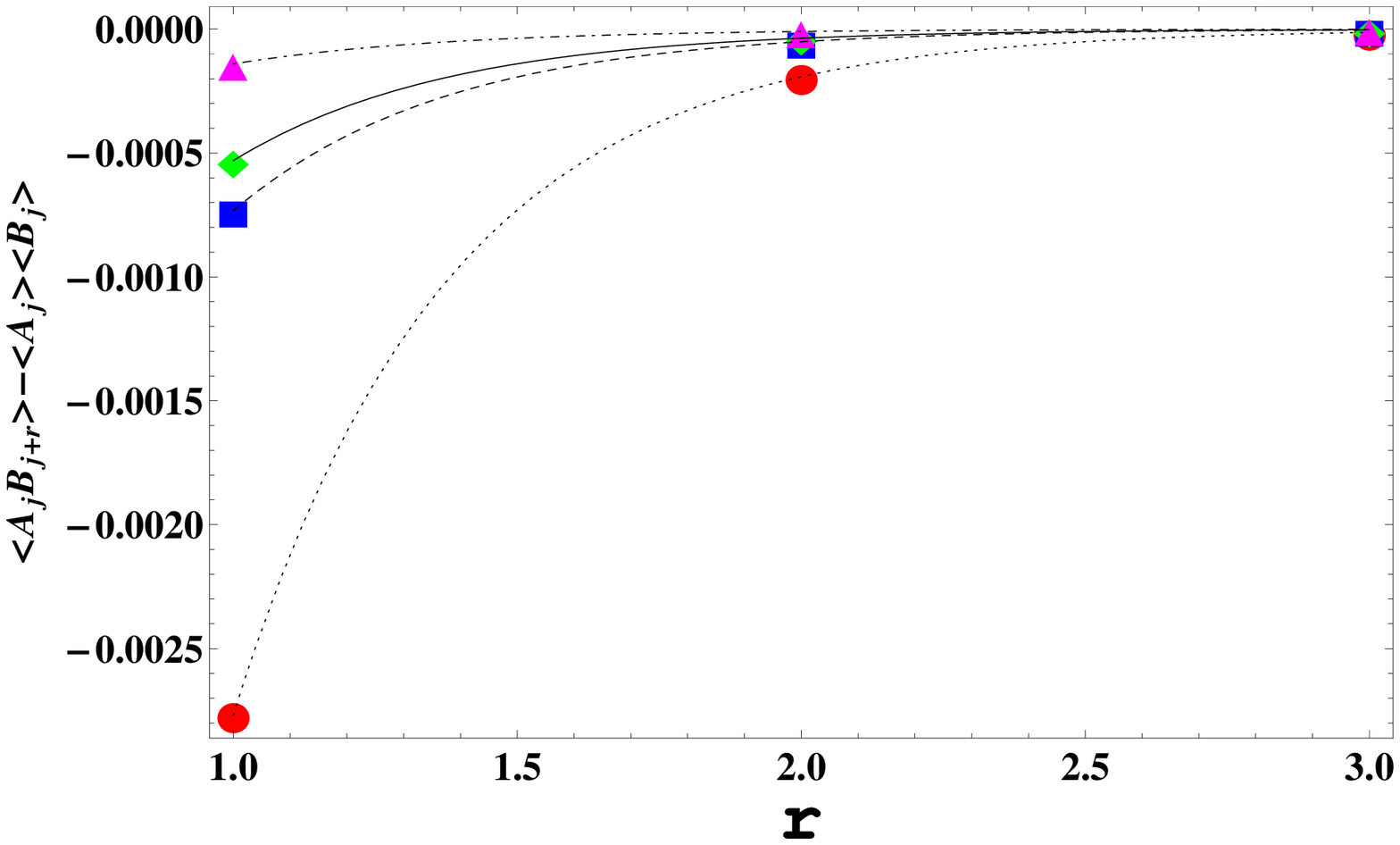}
\end{tabular}
\begin{tabular}{cc}
\includegraphics[width=\columnwidth]{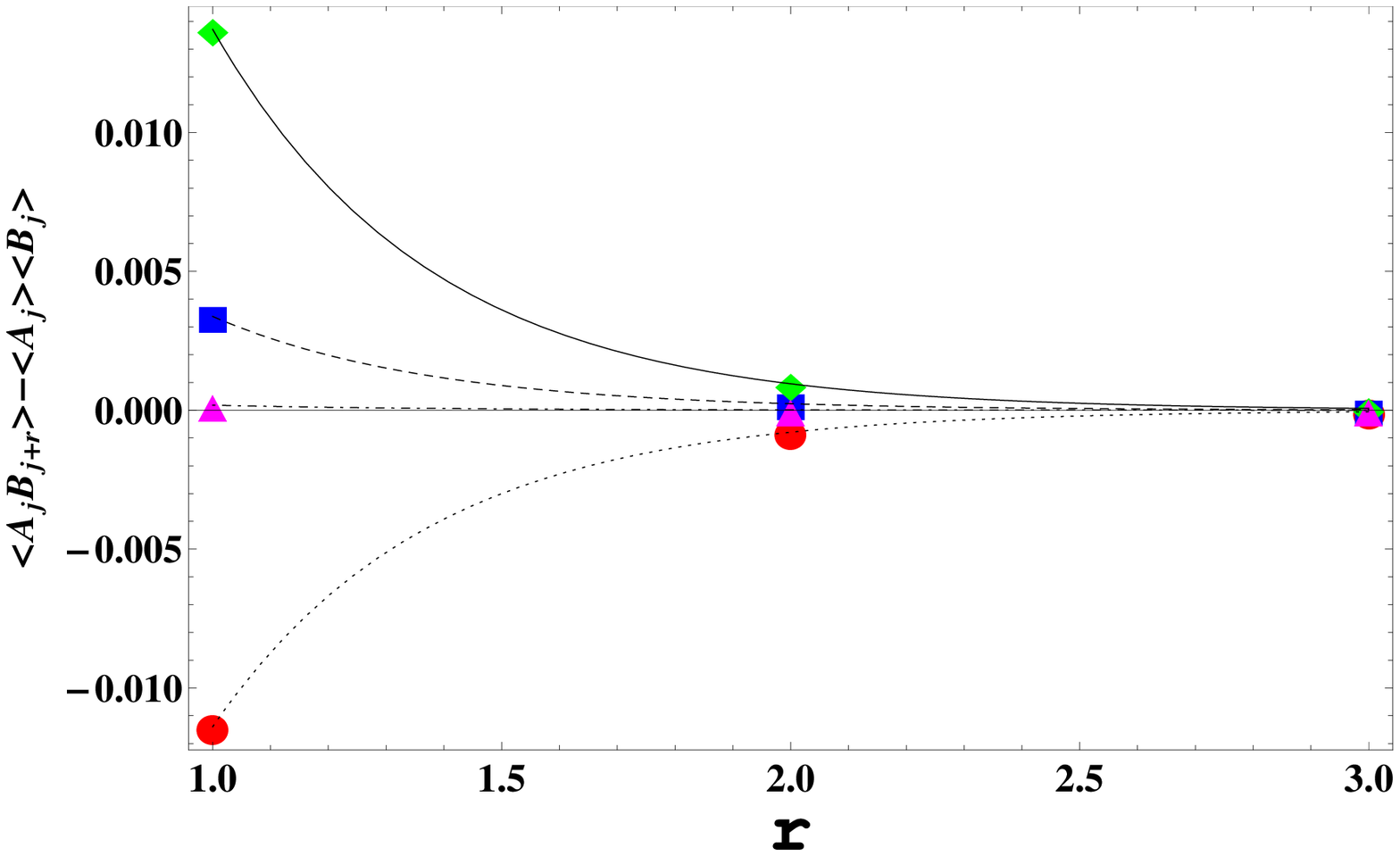}
\includegraphics[width=\columnwidth]{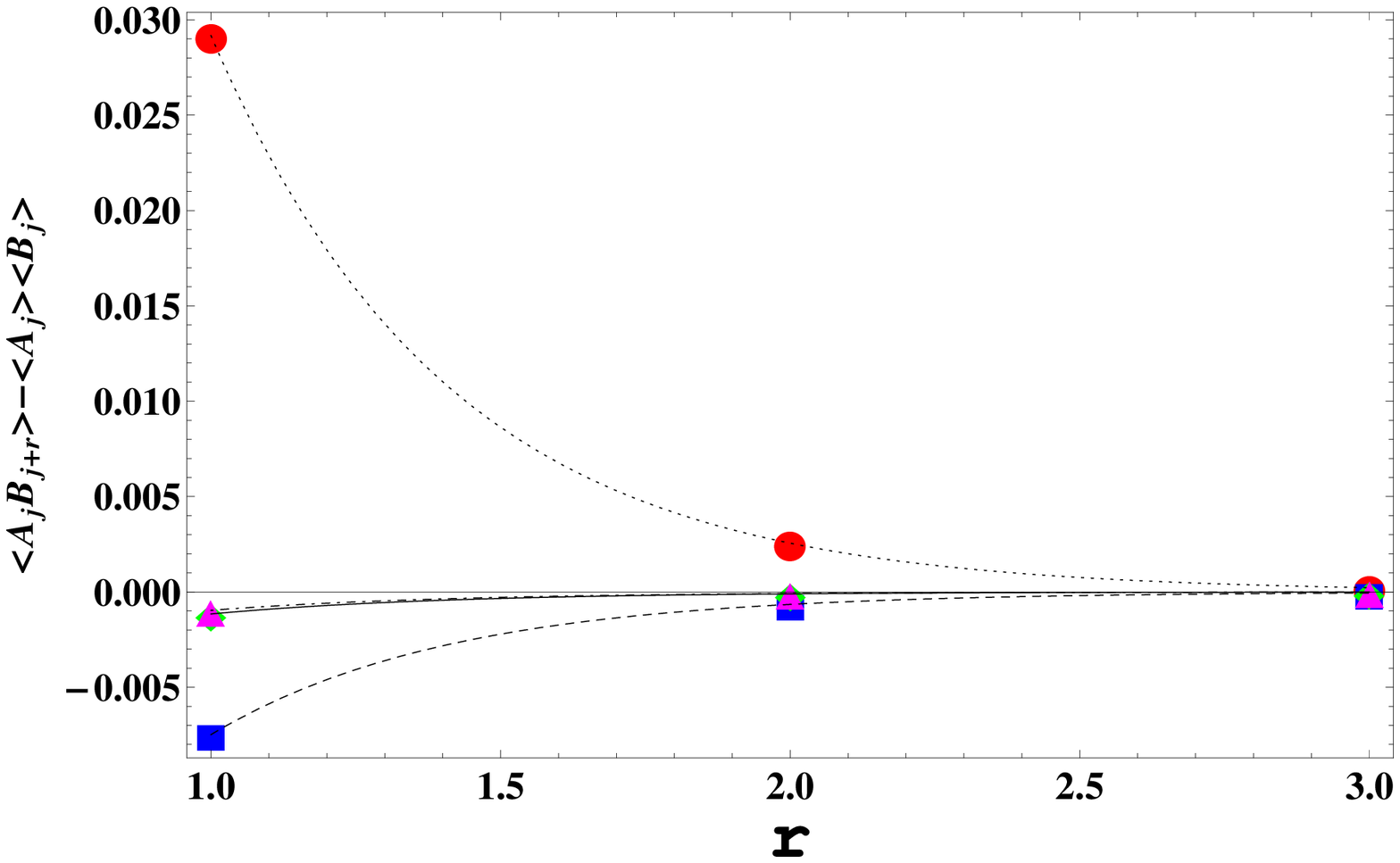}
\end{tabular}
\caption{\label{fig4}(Color online) The dependence of the
correlation functions for the asymmetric sawtooth chain against the
distance for $K/J=0.5, H/J=0.5, \Delta=0.3$ at $r=1$,$2$ and $3$
(See Eq. \ref{SC_all_corr}) The lines are just the guides for the
eyes. Upper left panel: $T/J=0.25$, $\langle S_{j,1}^z
S_{j+r,1}^z\rangle-\langle S_{j,1}^z\rangle^2$ - red circles
(dotted); $\langle S_{j,1}^z S_{j+r,2}^z\rangle-\langle
S_{j,1}^z\rangle\langle S_{j,2}^z\rangle$ - blue squares (dashed);
$\langle S_{j,2}^z S_{j+r,1}^z\rangle-\langle
S_{j,1}^z\rangle\langle S_{j,2}^z\rangle$ - green diamonds (solid);
$\langle S_{j,2}^z S_{j+r,2}^z\rangle-\langle S_{j,2}^z\rangle^2$ -
magenta triangles (dot-dashed). Upper right panel: $T/J=0.25$,
$\langle S_{j,1}^z \sigma_{j+r}\rangle-\langle
S_{j,1}^z\rangle\langle \sigma_j\rangle$ - red circles (dotted);
$\langle S_{j,2}^z \sigma_{j+r}\rangle-\langle
S_{j,2}^z\rangle\langle \sigma_j\rangle$ - blue squares (dashed);
$\langle S_{j,1}^z \tau_{j+r}\rangle-\langle S_{j,1}^z\rangle\langle
\tau_j\rangle$ - green diamonds (solid); $\langle S_{j,2}^z
\tau_{j+r}\rangle-\langle S_{j,2}^z\rangle\langle \tau_j\rangle$ -
magenta triangles (dot-dashed). Lower left panel: $T/J=0.25$,
$\langle \sigma_{j}S_{j+r,1}^z\rangle-\langle
S_{j,1}^z\rangle\langle \sigma_j\rangle$ - red circles (dotted),
$\langle \sigma_{j}S_{j+r,2}^z\rangle-\langle
S_{j,2}^z\rangle\langle \sigma_j\rangle$ - blue squares (dashed),
$\langle\sigma_{j}\sigma_{j+r,1}\rangle-\langle \sigma_j\rangle^2$ -
green diamonds (solid); $\langle
\sigma_{j}\tau_{j+r,1}\rangle-\langle \sigma_j\rangle\langle
\tau_j\rangle$ - magenta triangles (dot-dashed). Lower right panel:
$T/J=0.15$, $\langle  \tau_{j} S_{j+r,1}^z\rangle-\langle
S_{j,1}^z\rangle\langle \tau_j\rangle$ - red circles (dotted);
$\langle \tau_{j} S_{j+r,2}^z\rangle-\langle S_{j,2}^z\rangle\langle
\tau_j\rangle$ - blue squares (dashed); $\langle \tau_{j}
\sigma_{j+r}\rangle-\langle \sigma_{j}\rangle\langle \tau_j\rangle$
- green diamonds (solid); $\langle  \tau_{j}
\tau_{j+r}\rangle-\langle \tau_j\rangle^2$ - magenta triangles
(dot-dashed).}
\end{figure*}

\begin{figure*}[tbph]
\begin{tabular}{cc}
\includegraphics[width=\columnwidth]{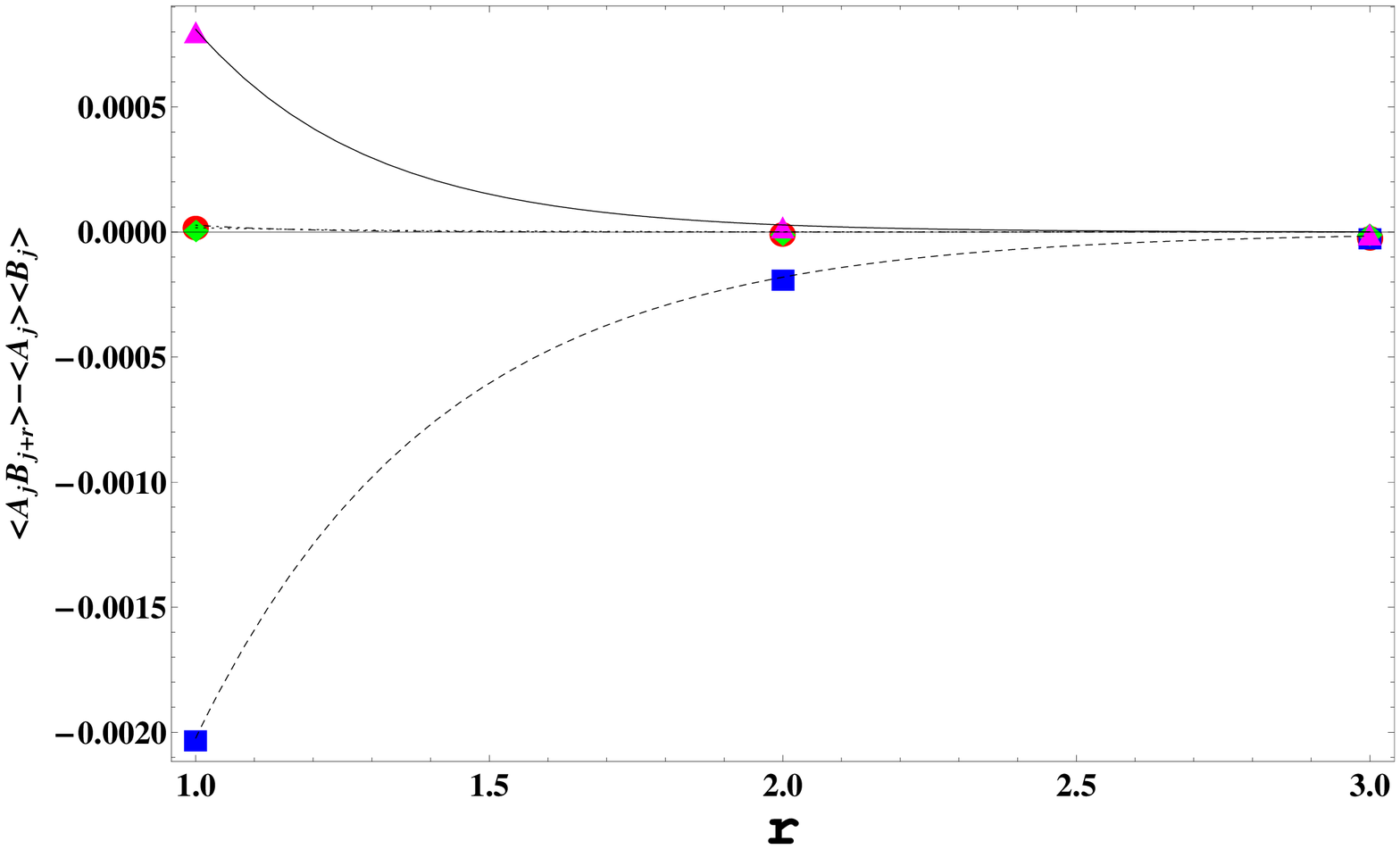}&
\includegraphics[width=\columnwidth]{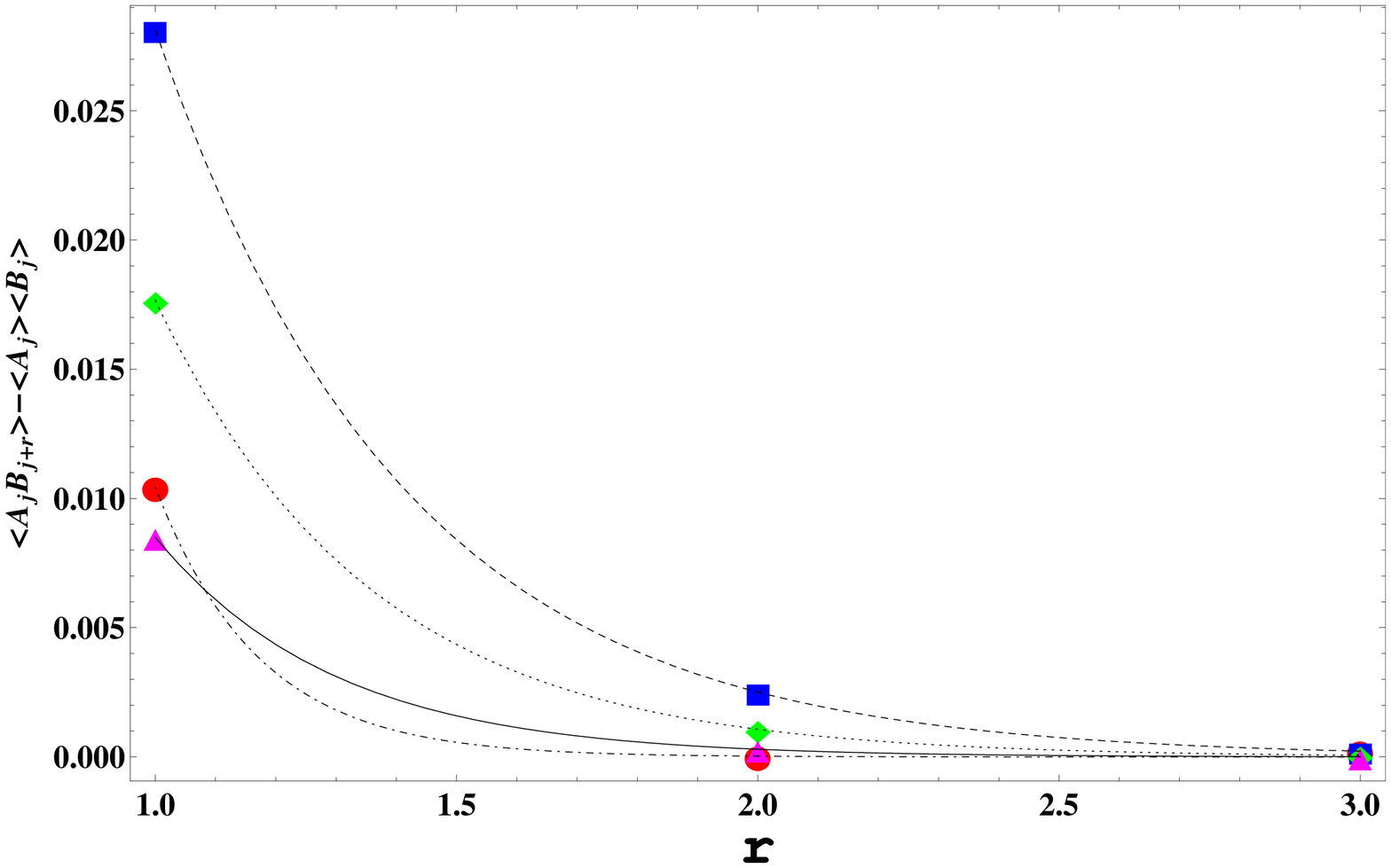}
\end{tabular}
\caption{\label{fig5}(Color online) The dependence of the
correlation functions $\langle S_{j,1}^z\tau_{j+r}\rangle-\langle
S_{j,1}^z\rangle\langle \tau_j\rangle$ (left panel) and $\langle
\tau_{j} S_{j+r,1}^z\rangle-\langle S_{j,1}^z\rangle\langle
\tau_j\rangle$ (right panel) against the distance for different
temperatures at $r=1$, $2$ and $3$. The lines are just the guides
for the eyes. $T/J=0.05$ - red bullets (dot-dashed ); $T/J=0.18$ -
blue squares (dashed); $T/J=0.28$ - green diamonds (dotted);
$T/J=0.42$ - magenta triangles (solid). }
\end{figure*}

 To illustrate the spatial and temperature dependence of various
correlation functions of Eq.(\ref{SC_all_corr}) it is helpful to
look at their plots. First of all, the general remarks about
correlations in HIC are noteworthy. The general structure of the HIC
exclude any possibility to have an entangled non trivial macroscopic
ground state, on the contrary, any macroscopic ground state is just
a tensor product of the local ground states, which, in their turn,
could be entangled. Thus, the zero-temperature long-range
correlations between the spins from different blocks, calculated
with respect to their expectation values, are trivial which implies
the zero value of all correlation function from Eq.
(\ref{SC_all_corr}) at $T=0$, as for the arbitrary ground state
$|\Psi\rangle=\prod_{i=1}^N |\psi_i\rangle$ one has
\begin{eqnarray}\label{triv_corr}
&&\langle \Psi|P_j R_{j+r} |\Psi\rangle=\langle
\psi_j|P_j|\psi_j\rangle\langle
\psi_{j+r}|R_{j+1}|\psi_{j+r}\rangle=\nonumber\\
&&\langle \Psi|P_j|\Psi\rangle\langle \Psi|R_{j+r}|\Psi\rangle.
\end{eqnarray}
 However, the thermal fluctuations as
well as the effect of the magnetic field lead to non-zero
correlations within a narrow interval of low temperatures for short
distances. The plots of the temperature dependence of all
correlation functions for the asymmetric sawtooth chain are
presented in Figure \ref{fig3} for the distance equal to one block,
$r=1$. Here, the values of the parameters correspond to the
spin-modulated zero-temperature ground state of the asymmetric
Heisenberg-Ising sawtooth chain\cite{oha09,bel10}. It is rather
expectable that the finite-temperature correlations are mostly of
short range with respect to blocks, but as one block contains four
spins, the non-zero, though weak, correlations even for $r=1$ are
the non-trivial result. One can see from the plots in Figure
\ref{fig3} the vanishing of all correlations at high as well as at
low temperatures as was mentioned above. As expected, the closer to
each other are situated the spins from different blocks, the larger
is the magnitude of the correlation between them. For instance, the
correlation between $\tau_j$ and $S_{j+1,1}^z$ at the temperature
corresponding to the peak is four times larger in absolute value
than the correlation between $\tau_j$ and $S_{j+1,2}^z$ and more
that seventy times larger than the correlation between $S_{j,2}^z$
and $\tau_{j+1}$. In the same time, the correlation between
$S_{j,1}^z$ are the strongest ones, though, the are separated by The
dependence of the correlation functions against the distance for the
same values of coupling constants and anisotropy and for
temperatures corresponding to the strongest correlation of the
thermal fluctuations are presented in Figure \ref{fig4}. For all
correlation functions one can see the usual exponential decay in
their absolute value. The sign of the correlations is defined by the
mutual orientation of the corresponding spins which can be seen in
the $T=0$ ground state\cite{oha09,bel10} for the given set of the
parameters. In Figure \ref{fig5} we presented the plots of the same
correlation functions  $\langle  S_{j,1}^z \tau_{j+r}\rangle-\langle
S_{j,1}^z\rangle\langle \tau_j\rangle$ and $\langle \tau_{j}
S_{j+r,1}^z\rangle-\langle S_{j,1}^z\rangle\langle \tau_j\rangle$
against the distance for different temperatures. The non-monotonous
behavior of the correlation functions described above can be clearly
seen from the plots.

\section{Conclusion}
In the present paper we elaborated a technique for calculating
correlation functions in HIC. The technique is based on the
classical transfer-matrix formalism which proves itself as the most
efficient and straightforward method of treating the HIC, allowing
one to obtain exact expressions for the partition function and,
thus, for all thermodynamic functions in a quite simple
way\cite{val08,ant09,oha09, oha10, bel10, roj11a, roj11b,
lis11a,lis11b, oha12,cha12,roj12a,ana12,roj12b}. However, the
classical transfer-matrix formalism is also very convenient for the
calculation of the correlation functions of various kinds, because
all canonical expectation values are expressed in terms of traces of
certain products of the transfer-matrices with matrices representing
the microscopic variables in the transfer-matrix
formalism\cite{bax}. We presented a detailed scheme of calculation
of the correlation functions for HIC, obtained general expressions
for the corresponding trace formulas and presented the general
expressions for the matrix representations of the main microscopic
variables for the typical HIC, quantum spin operators, additional
classical spins from block Hamiltonians and intermediate spin. As it
is expected from general principles, the spatial decay of any
correlation between any pair of spin is characterized by a single
correlation length obtained in a standard way as the reciprocal of
the logarithm of the ratio of the largest and second largest
eigenvalues of the classical transfer matrix. In the simplest case
considered here when intermediate classical $\sigma$-spins are 1/2
spins the transfer matrix has dimension 2 by 2 and, thus, only two
eigenvalues. However, the prefactors are strongly affected by the
microscopic variables. We illustrated our general formalism by two
examples, the simplest HIC, symmetric Heisenberg-Ising diamond-chain
and asymmetric Heisenberg-Ising sawtooth chain with a pair of
quantum spins in the unit cell. We presented the exact form of the
transfer-matrix, its eigenvalues, matrix representation
corresponding to the classical transfer-matrix formalism for quantum
spin operators, matrix representation for additional classical spin
from block Hamiltonian and all possible pair correlation functions.
A very important feature of the correlations function of the HIC
considered in the present paper is the absence of correlations
between transverse components of quantum spins from different
blocks. The corresponding matrix representation of the $S_{n,1}^x$
and $S_{n,a}^y$ operators are identically zero. This fact is an
obvious consequence of the structure of the HIC where blocks
interact to each other only through $z$-components of the spins.

It is also worth mentioning that the general idea of using the
classical transfer-matrix for describing the thermodynamic
properties of decorated spin lattices can be applied also for
two-dimensional systems. Although the existing examples of the
papers devoted to the topic deal with the decoration-iteration
transformation\cite{fisher,str10c,2d1,2d2,2d3,2d4}, the direct
generalization of the existing transfer-matrix method is possible.
As known, two-dimensional systems can exhibit phase transitions at
finite temperature in contrast to the one-dimensional models where
only zero-temperature transitions are possible. In principle,
two-dimensional spin systems can possess a complicated structure of
the order parameters, and multiple orderings, which, particularly,
can be different in different spatial directions. For these reasons,
the statement about the unique correlation length is no longer valid
here.

\section{Acknowledgements}
V.O. expresses his gratitude to the LNF-INFN in Frascati for warm
hospitality during the work on the paper. He also acknowledges
partial financial support from Volks\-wagen Foundation (Grant No.\
I/84 496) and from the project SCS-BFBR 11RB-001.

\end{document}